\PassOptionsToPackage{sort&compress}{natbib}
\documentclass[final,5p]{elsarticlev33}
\usepackage{amsmath,amssymb,amsfonts,amsthm,makeidx,graphicx,booktabs}
\usepackage{cuted}
\usepackage{fancyhdr}
\usepackage{newtxtext,newtxmath}
\usepackage{flushend}
\usepackage{stfloats}

\usepackage{etoolbox}
\AtBeginEnvironment{thebibliography}{\sffamily}

\usepackage{xcolor}
\definecolor{myredbrown}{rgb}{0.55, 0.0, 0.0}
\definecolor{myblue}{rgb}{0.1, 0.2, 0.6}
\usepackage[colorlinks=true]{hyperref}
\AtBeginDocument{%
  \hypersetup{
    linkcolor = myblue, 
    citecolor = myredbrown, 
    urlcolor  = myblue      
  }%
}

\fancypagestyle{elsevierhead}{%
  \fancyhf{}%
  \fancyhead[R]{Neutrinos from core-collapse supernovae}%
  \fancyhead[L]{G.~G.~Raffelt, H.-T.~Janka, D.~F.~G.~Fiorillo}%
  \fancyfoot[C]{\thepage}%
}

\usepackage{microtype}
\makeatletter
\def\@textbottom{\vskip \z@ \@plus 1pt}
\let\@texttop\relax
\makeatother
\makeatletter
\fancypagestyle{pprintTitle}{%
  \fancyhf{}%
  \fancyhead[R]{Neutrinos from core-collapse supernovae}%
  \fancyhead[L]{Raffelt, Janka, Fiorillo}%
  \fancyfoot[C]{\thepage}%
}
\makeatother

\makeatletter
\providecommand\@combinedblfloats{}
\providecommand\@setmarks{}
\makeatother

\newcommand{\bp}{{\bf p}}
\newcommand{\bv}{{\bf v}}
\newcommand{\br}{{\bf r}}
\newcommand{\bK}{{\bf K}}
\newcommand{\GF}{G_{\rm F}}

\newcommand{\TCH}[1]{{\em #1}}
\long\def\exclude#1{}

\usepackage{caption}
\usepackage{tocloft}
\begin{document}

\begin{frontmatter}

\title{\textsf{\textbf{Neutrinos from core-collapse supernovae}}
}

\tnotetext[t1]{To be published in {\em Encyclopedia of Particle Physics} (Elsevier)}

\author[AddressA]{Georg G.~Raffelt}
\author[AddressB]{Hans-Thomas Janka}
\author[AddressC]{Damiano F.~G.~Fiorillo}

\address[AddressA]{Max-Planck-Institut f\"ur Physik, Boltzmannstr.~8, 85748 Garching, Germany}
\address[AddressB]{Max-Planck-Institut f\"ur Astrophysik, Karl-Schwarzschild-Str.~1, 85748 Garching, Germany}
\address[AddressC]{Deutsches Elektronen-Synchrotron DESY, Platanenallee 6, 15738 Zeuthen, Germany}

\begin{abstract}
The core of a massive star ($M\gtrsim8\,M_\odot$) eventually collapses. This implosion usually triggers a supernova (SN) explosion that ejects most of the stellar envelope and leaves behind a neutron star (NS) with a mass of up to about $2\,M_\odot$. Sometimes the explosion fails and a black hole forms instead. The NS radiates its immense binding energy (some 10\% of its rest mass or 2--$4\times10^{53}~{\rm erg}$) almost entirely as neutrinos and antineutrinos of all flavors with typical energies of some 10~MeV. This makes core-collapse SNe the most powerful neutrino factories in the Universe. Such a signal was observed once---with limited statistics---from SN~1987A in the Large Magellanic Cloud. Today, however, many large neutrino detectors act as SN observatories and would register a high-statistics signal. A future Galactic SN, though rare (1--3 per century), would produce a wealth of astrophysical and particle-physics information, including possible signatures for new particles. Neutrinos are key to SN dynamics in the framework of the Bethe--Wilson delayed explosion paradigm. After collapse, they are trapped in the core for a few seconds, forming a dense neutrino plasma that can exhibit collective flavor evolution caused by the weak interaction, a subject of intense theoretical research.
\smallskip
\end{abstract}


\end{frontmatter}

\thispagestyle{empty}

\tableofcontents

\newpage

\pagestyle{elsevierhead}

\section{Introduction}\label{sec:introduction}

\noindent Collapsing stars are the most powerful astrophysical~sources for low-energy (MeV-range) neutrinos and release, within a few seconds, an amount of energy that is equivalent to some 10\% of the rest mass of the neutron star (NS) that forms in the implosion. In usual astrophysical units, this is an energy of 2--$4\times10^{53}~{\rm erg}$, or during the emission of a few seconds, around $10^{53}~{\rm erg}~{\rm s}^{-1}$. This neutrino luminosity corresponds to a few $10^{19}\,L_\odot$ (solar luminosities, $L_\odot=3.83\times10^{33}\,{\rm erg}~{\rm s}^{-1}$). Thus, for a few seconds, the supernova (SN) shines more brightly in neutrinos than the rest of the Universe shines in photons. A core-collapse SN happens around once every second in the visible Universe, implying that stars radiate as much energy in SN neutrinos as in photons. Therefore, the energy density of the cosmic diffuse SN neutrino background (DSNB) is comparable to the extragalactic background light, the electromagnetic energy emitted by all stars. The DSNB is the subject of another chapter in this {\em Encyclopedia}; here, we review the physics of core-collapse SNe and the concomitant role of neutrinos.

The core of a collapsed massive star reaches densities beyond that of nuclei ($\rho_{\rm nuc}\simeq3\times10^{14}~{\rm g}~{\rm cm}^{-3}$), trapping even neutrinos---that reach energies beyond 100~MeV---despite their feeble interactions. SN cores are among the few classes of astrophysical environments where neutrinos reach thermal equilibrium. Other examples are NS mergers (covered in another chapter of this {\em Encyclopedia}) as well as the early universe (covered in yet another~chapter).

The geometric dimension of a SN core is tens of km, whereas a typical neutrino mean free path is perhaps a meter, but strongly depends on energy, flavor, and the density and chemical composition of the nuclear medium. On the other hand, photons interact much more strongly, are essentially stuck, and therefore cannot transport energy efficiently. The capability of diffusive energy transport is directly proportional to the mean free path of a given form of radiation. Of course, once it exceeds the geometric dimensions of the star, it simply carries away energy instead of transporting it between different regions. An example is gravitational radiation in the form of gravitons, which interact so feebly that even nuclear matter is transparent to them and they carry away only a minor fraction of the SN energy. 

\begin{figure}
\vskip4pt
	\centering
	\hbox to\columnwidth{\includegraphics[width=0.49\columnwidth]{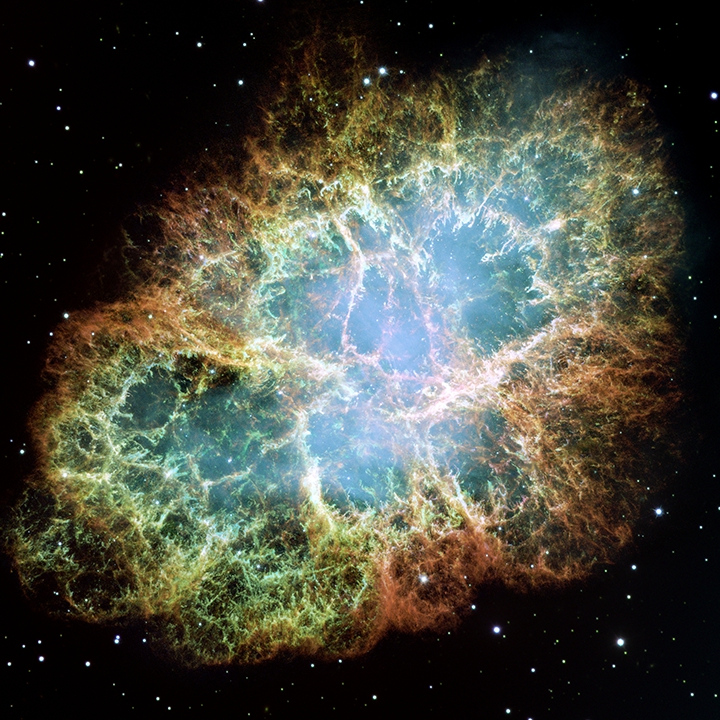}\hfil
    \includegraphics[width=0.49\columnwidth]{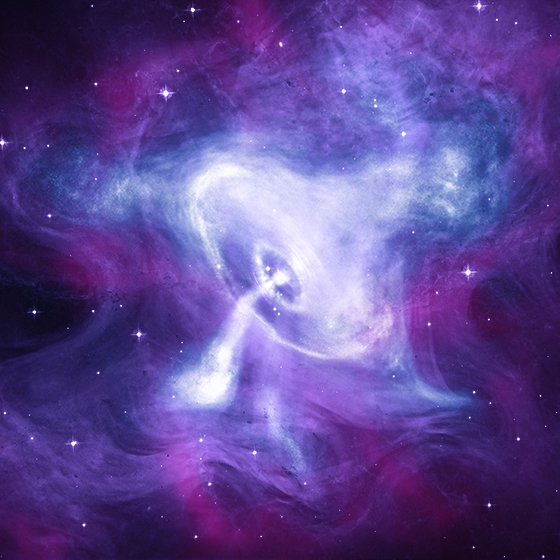}}
	\caption{Historical supernova of 1054 that has produced the Crab Nebula (left) and a compact remnant, the Crab Pulsar (right), a pulsating neutron star, near the center of the nebula. The gravitational and nuclear binding energy of the neutron star, some 10\% of its rest mass, is emitted as a neutrino burst that lasts for a few seconds and drives the explosion. Such a burst was observed only once (from the historical SN~1987A on 23 February 1987). A future nearby supernova could be seen with high statistics in many active neutrino detectors and possibly in gravitational waves. The Crab Nebula image was taken by the Hubble Space Telescope (credit \href{https://hubblesite.org/contents/media/images/3885-Image}{NASA/STScI}). The main pulsar image was taken in X-rays by the Chandra satellite, superimposed with an optical (HST) and infrared (Spitzer) image (credit \href{https://chandra.harvard.edu/photo/2018/crab/}{NASA/Chandra}).}
	\label{fig:titlepage}
\end{figure}

Overall, neutrinos interact strongly enough to be trapped, yet feebly enough to have a mean free path not much smaller than the geometric dimensions of a SN core. Therefore, they are the main carriers of energy and lepton number.

They also drive the explosion dynamics, influence nucleosynthesis, and kick the NS and produce gravitational waves (GWs) when emitted anisotropically. Moreover, they can be detected at Earth, allowing one to probe SN physics with neutrinos as astrophysical messengers. A future high-statistics SN neutrino observation would provide a wealth of astrophysical and particle-physics information, although Galactic SNe unfortunately occur only 1--3 times per century. A first teaser was provided by the historical SN~1987A in the Large Magellanic Cloud (LMC), a small satellite galaxy of our Milky Way, whose first signal included two dozen neutrinos observed on 23~February~1987. The time scale of a few seconds and the energies of a few 10~MeV corresponded to expectations. 

In addition, hypothetical low-mass particles, notably axions or general Feebly Interacting Particles (FIPs), can be produced in the core and modify SN physics. Therefore, core-collapse SNe also serve as laboratories to learn about such particles or nonstandard neutrino properties.

There exist excellent reviews of the SN explosion mechanism from a modern perspective \cite{Burrows:2000mk, Woosley:2002zz, Janka:2006fh, Janka:2012wk, Muller:2016izw, Muller:2019upo, Burrows:2019zce, Muller:2020ard, Burrows:2020qrp, Boccioli:2024abp, Janka:2025tvf, Jerkstrand:2025}, and of the neutrino signal, flavor oscillations, and detection opportunities~\cite{Scholberg:2012id, Mirizzi:2015eza, Volpe:2023met, Vartanyan:2023zlb, Choi:2025igp, Tamborra:2024fcd, SNEWS:2020tbu}. Reviews on any aspect of SN physics are available in the Handbook of Supernovae~\cite{Alsabti:2017ahu}. We will reference various specific results, but we will not attempt an exhaustive bibliography and instead refer the reader to this extensive review literature. Following particle-physics practice, we will use natural units for most formulas, where \hbox{$\hbar=c=k_{\rm B}=1$}.

\section{Stellar collapse and explosion}
\label{sec:Explosion}

\subsection{Neutrino driven supernova explosion}

\noindent Stars condense from clouds of interstellar matter, consisting mostly of hydrogen, a quarter by mass of helium that was mainly produced in the early universe, and a small fraction of heavier elements (``metals'') that originate from previous generations of stars. After contraction, the proto star eventually ignites the burning of hydrogen at its center and evolves as a ``main sequence star'' like our Sun until hydrogen is exhausted at its center. This helium core later ignites fusion to carbon, while hydrogen keeps burning in a shell. 

Stars with masses below about $8\,M_\odot$ never get hot and dense enough to ignite further phases of nuclear burning. Instead, they shed their envelopes and leave a compact remnant, a carbon-oxygen white dwarf (WD) with a typical mass of around $0.6\,M_\odot$, although the distribution is broad. For this specific mass, the radius is some 8,000~km---or roughly the size of the Earth---and famously smaller for larger masses. A WD is supported by electron degeneracy and therefore stable without thermal pressure; it does not need nuclear burning to avoid collapse as long as its mass is below the Chandrasekhar limit of around $1.4\,M_\odot$, the largest mass that can be supported by degenerate electrons. After formation, it simply cools until it fades away. 

Occasionally, it may later accrete matter from a binary companion. If nuclear fusion is triggered in the compression-heated accretion layer or WD interior, a nuclear run-away and stellar explosion may occur. Such thermonuclear SN explosions (spectral type Ia, defined by the absence of hydrogen and helium lines and the presence of silicon lines) leave an extended nebula, but in most cases no compact remnant. The mass-donating companion is probably a helium star without a hydrogen envelope. Alternatively, in a perhaps more frequent path, such a SN results from the final interaction or possibly merger of two WDs in a binary system \cite{Ruiter:2024kak}.
Assuming nuclear fusion of $1.0\,M_\odot$ ($1.2\times 10^{57}$ nucleons) and an available nuclear binding energy of 1~MeV/nucleon, a total energy of $2\times10^{51}~{\rm erg}$ is released. Around one third of all SN explosions are of the thermonuclear type (spectral type Ia).

\begin{figure*}
\vskip12pt
    \centering
    \includegraphics[width=0.70\textwidth]{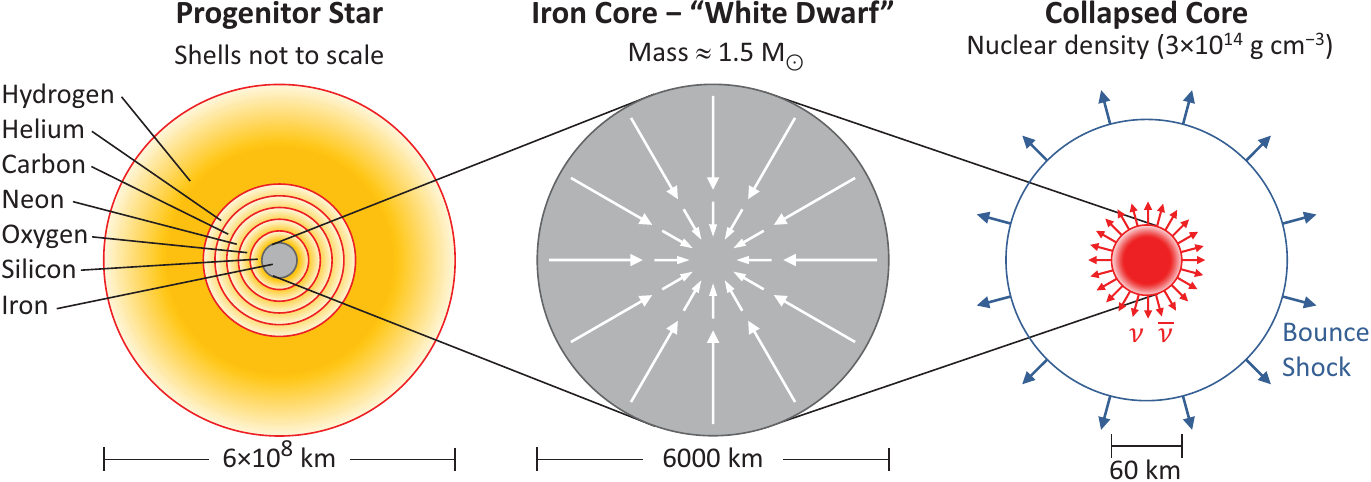}
    \caption{Core collapse and subsequent supernova (SN) explosion of a massive star ($M\gtrsim 8\,M_\odot$) as described in the text. The collapsed SN core, forming a proto-neutron star, is so dense and hot that the thermal neutrinos are trapped and carry away the gravitational binding energy of \hbox{$2$--$4\times10^{53}~{\rm erg}$} of the final neutron star over a period of several seconds.}
     \label{fig:SN-Collapse}
\end{figure*}

Stars more massive than $8\,M_\odot$ follow a different path in that they pass through all nuclear burning phases, eventually forming the famous onion-skin structure of different burning shells sketched in Fig.~\ref{fig:SN-Collapse}. After the end of core hydrogen burning, stars of any mass expand their envelopes to become hugely extended giant stars, whereas the inner structure remains more compact. For a $15\,M_\odot$ progenitor as an example, core hydrogen burning would last about $11\times10^6$ years, core burning of helium $2\times10^6$~y, carbon 2,000~y, neon 0.7~y, oxygen 2.6~y, and finally silicon up to three weeks \cite{Woosley+2005}. At this point, the star has developed a compact iron core, essentially an iron WD supported by electron degeneracy pressure. Iron is the most tightly bound nucleus and no further nuclear energy can be released by fusion reactions. As silicon burning proceeds, the iron core grows in mass and shrinks in size, following the inverse mass-radius relation ($R\propto M^{-1/3}$) characteristic of degenerate stars. Eventually, close to the Chandrasekhar limit, photodisintegration and electron captures of (iron-group) nuclei set in and destabilize the stellar core with the inevitable consequence of gravitational collapse.

The collapse continues until the core reaches supra-nuclear density and the equation of state stiffens, suddenly halting the implosion. This core bounce spawns a shock wave that moves outward, eventually exploding the outer layers, i.e., the implosion is turned around, while leaving behind the collapsed core, a proto-neutron star (PNS) or nascent neutron star, which cools by neutrino emission until it settles to a compact radius of 12--14~km and a mass of up to around $2\,M_\odot$. 

The gravitational binding energy liberated by compressing this mass to such a small size is some 10\% of its rest mass, corresponding to around 100~MeV per nucleon. However, 99\% of this energy is carried away by neutrinos so that only around 1~MeV/nucleon remains for the explosion and photon luminosity, comparable to the kinetic energy of a thermonuclear SN. In both cases, the scale for the explosion energy is $10^{51}~{\rm erg}$, sometimes called 1~foe (ten to ``fifty-one ergs'') or nowadays called 1~bethe (1\,B). The nuclear processed material ejected in both classes of SNe accounts for practically all elements heavier than helium in the universe, which are subsequently available for the formation of the next generation of stars and their planets.

\begin{figure}[b!]
    \centering
    \includegraphics[width=0.999\columnwidth]{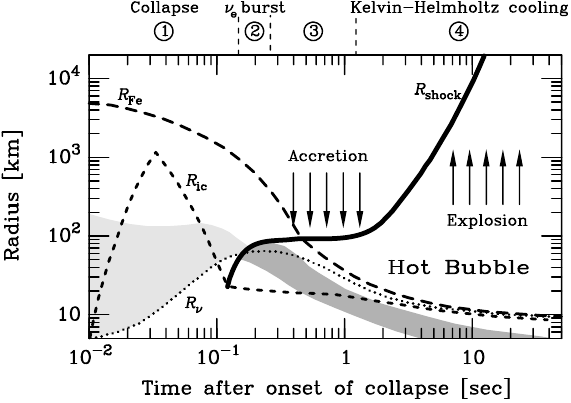}
    \caption{Schematic picture of the core collapse of a massive star ($M\gtrsim 8\,M_\odot$), of the formation of a neutron-star (NS) remnant, and the beginning of a supernova (SN) explosion. There are four main phases numbered 1--4 above the plot: (1)~Collapse. (2)~Prompt-shock propagation and break-out through the neutrino sphere, release of prompt $\nu_e$ burst. (3)~Matter accretion and mantle cooling. (4)~Kelvin-Helmholtz cooling of proto-neutron star (PNS). The curves mark the evolution of several characteristic radii: The stellar iron core ($R_{\rm Fe}$). The neutrino sphere ($R_\nu$) with diffusive transport inside, free streaming outside. The inner core ($R_{\rm ic}$), which for $t\lesssim 0.1~{\rm s}$ is the region of subsonic collapse, later it is the settled, compact inner region of the nascent neutron star. The SN shock wave ($R_{\rm shock}$) is formed at core bounce, stagnates for up to several $100~{\rm ms}$, and is revived by neutrino heating---it then propagates outward and ejects the stellar envelope. The shaded area is where most of the neutrino emission comes from; between this area and $R_\nu$, neutrinos still diffuse, but are no longer efficiently produced. (Adapted from \cite{Raffelt:1996wa}, \copyright~University of Chicago Press. Figure originally adapted from Janka~\cite{Janka:1992jk}.) }
     \label{fig:Collapse-Scheme}
\end{figure}

\begin{figure*}
\vskip6pt
    \centering
    \includegraphics[width=1.0\textwidth]{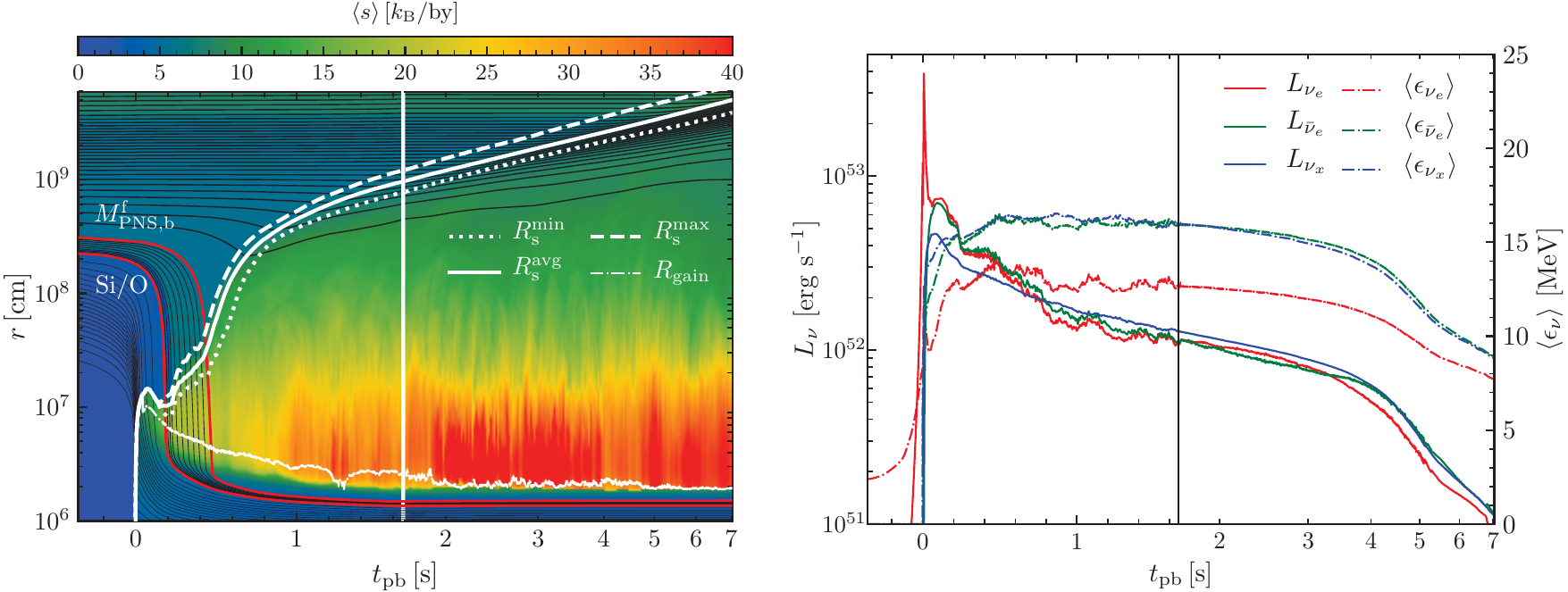}
    \caption{Three-dimensional simulation by the Garching group of a $19\,M_\odot$ progenitor \cite{Bollig:2020phc}. Up to the vertical line, the postbounce (pb) time axis is linear, beyond it is logarithmic, and up to that break, neutrino transport was done with the {\sc Vertex} code, whereas for the long-term evolution, a numerically cheaper heating and cooling scheme was applied. {\bf Left.} Mass shells (black lines) and entropy per nucleon color-coded. Maximum, minimum, and average shock radii as white lines, the gain radius is the lowest white line, and in red, the mass shells of Si/O shell interface and final neutron-star mass boundary. {\bf Right.} Neutrino luminosities and mean energies of $\nu_e$, $\overline\nu_e$, and one species $\nu_x$ of heavy-lepton neutrinos $\nu_\mu$, $\overline\nu_\mu$, $\nu_\tau$, or $\overline\nu_\tau$.    (Figure from \cite{Bollig:2020phc} with permission of the AAS.) 
    }
    \label{fig:SN-3D}
\end{figure*}

The core-collapse explosion mechanism is actually a bit more complicated because the bounce shock loses too much energy before it could drive what would be a prompt explosion. The shock forms within the iron core and on its way out dissociates iron nuclei into free nucleons, loses energy, stalls, and eventually retreats, while material keeps falling in, leading to black-hole (BH) formation. Indeed, this is what probably happens in up to 25\% of all cases: the class of failed SNe. 

While the shock wave lingers at a radius of 150--200~km, material keeps falling in and powers neutrino emission from the outer PNS region. As they stream out, neutrinos deposit some of their energy by beta processes of the type $\nu_e+n\to p+e^-$ and $\overline\nu_e+p\to n+e^+$ in a region below the shock called gain region, where these reactions dominate over the inverse ones. Between the gain radius and the shock wave, the net effect of neutrinos is energy deposition, heating the material, thus supplying renewed hydrodynamical pressure. After this accretion phase of up to a few 100~ms, the rejuvenated shock wave resumes moving toward the final explosion. 

This \textit{delayed explosion mechanism} was first proposed in 1985 by Bethe and Wilson \cite{Bethe:1985sox} and therefore is also called \textit{Bethe--Wilson--mechanism}, although the crucial role of neutrinos for initiating the explosion had been surmised already in the 1960s \cite{ColgateWhite1966, Arnett1967}. The decisive support by multi-dimensional fluid flows in the postshock region was realized only after SN~1987A (see Sec.~\ref{sec:non-spherical}), which showed strong deviations from spherical symmetry and radial mixing of chemical elements between core and envelope. This complicated sequence of events is sketched in Fig.~\ref{fig:Collapse-Scheme} with a description of the different phases.

The same evolution is shown in Fig.~\ref{fig:SN-3D} from a modern numerical simulation that was performed without the assumption of spherical symmetry, to be elaborated further in Sec.~\ref{sec:non-spherical}. Following the solid white line as a function of time post bounce (pb), we see how the shock wave forms and moves to some 150~km and then retracts, while mass accretes as we see from the mass shells (thin black lines) crossing the shock. When the Si/O interface passes at around 200~ms pb, the shock resumes moving and then keeps expanding, while some mass still falls in. The outer red line marks the mass coordinate of what becomes the final NS surface.

\subsection{Flavor dependent neutrino interactions and transport}

\noindent The six species $\nu_e$, $\overline\nu_e$, $\nu_\mu$, $\overline\nu_\mu$, $\nu_\tau$, and $\overline\nu_\tau$ play rather different roles because of the different masses of their charged-lepton partners. The electron mass $m_e=0.511~{\rm MeV}$ is much smaller than the typical 30--100~MeV energies in a SN core and actually, the electron refractive mass, caused by coherent interaction with the background medium, is much larger (around 10~MeV), so $m_e$ itself is negligible. The mass $m_\mu=105.7~{\rm MeV}$ of muons is small enough for them to be produced, but their impact is subdominant, whereas finally $m_\tau=1777\,{\rm MeV}$ (almost twice the nucleon mass) is so large that $\tau$ leptons never appear. For $\nu_e$ and $\overline\nu_e$, charged-current nuclear processes of the type $\nu_en\leftrightarrow pe^-$ and $\overline\nu_ep\leftrightarrow ne^+$ are the dominant interaction channels and main source and sink terms for their production and absorption. For $\nu_\mu$ and $\overline\nu_\mu$, analogous processes with muons occur, but suffer from the $m_\mu$ threshold~\cite{Bollig:2017lki, Bollig2018, Guo:2020tgx}. All heavy-flavor species can be produced in pairs by $e^+e^-\leftrightarrow \nu\overline\nu$ or by nuclear bremsstrahlung $NN\leftrightarrow NN\nu\overline\nu$. Their main interaction is neutral-current elastic scattering $\nu N\to N\nu$. 

A reference process for weak interactions is inverse beta decay (IBD) with a cross section, ignoring recoil effects, the electron mass, and the $p$--$n$ mass difference, of
\begin{eqnarray}\label{eq:sigma_CC}
    \sigma_{\bar\nu_ep\to n e^+}&\simeq&\frac{G_{\rm F}^2\,\cos^2\theta_{\rm C}}{\pi}
    \left(C_V^2+3C_A^2\right)\epsilon_\nu^2
    \nonumber\\
    &=&9.4\times10^{-40}~{\rm cm}^2\,\left(\frac{\epsilon_\nu}{100\,{\rm MeV}}\right)^2,
\end{eqnarray}
where $G_{\rm F}=1.166\times10^{-5}~{\rm GeV}^{-2}$ is the Fermi constant, $\cos^2\theta_{\rm C}=0.949$ refers to the Cabibbo angle, $C_V=1$ is the charged vector current constant, and $C_A=1.275$ the axial-current coupling constant \cite{ParticleDataGroup:2024cfk}. An accurate approximation formula for the IBD cross section is also available \cite{Strumia:2003zx}. 

For elastic neutrino-nucleon scattering, the corresponding neutral-current cross section is, with the same simplifications,
\begin{eqnarray}\label{eq:sigma_NC}
    \sigma_{\nu N\to N \nu}&=&\frac{G_{\rm F}^2}{4\pi}
    \left(C_V^2+3C_A^2\right)\epsilon_\nu^2
    \nonumber\\
    &\simeq& 
   \left.\begin{matrix} 2.1~\\ 2.5~ \end{matrix}\right\}
    \times10^{-40}~{\rm cm}^2\,\left(\frac{\epsilon_\nu}{100\,{\rm MeV}}\right)^2,
\end{eqnarray}
where the upper line refers to protons, the lower one to neutrons because $C_V=-1$ for neutrons and $1-4\sin^2\theta_{\rm W}\simeq0.075$ for protons with $\theta_{\rm W}$ the weak mixing angle. In both cases, $\sigma$ is dominated by the axial coupling, i.e., the spin-dependent piece of the weak interaction. In a SN core, these cross sections receive large corrections from degeneracy, the effective nucleon mass in a nuclear medium, many-body correlations, and a modification of the axial coupling constant. Even in vacuum, there are recoil and weak-magnetism corrections, and for the neutral-current case, $C_A$ differs from 1.27 by a strange-quark contribution. (See the review of measurements in Ref.~\cite{KamLAND:2022ptk} and Ref.~\cite{Melson:2015spa} for an implementation in a SN model.) 

All of these modifications cause numerical corrections, but not order-of-magnitude changes, so that Eqs.~\eqref{eq:sigma_CC} and \eqref{eq:sigma_NC} provide representative scales. Nuclear density of $3\times10^{14}~{\rm g}~{\rm cm}^{-3}$ corresponds to a baryon density of $n_0=0.181~{\rm fm}^{-3}$, where $1~{\rm fm}=10^{-13}~{\rm cm}$. For this density, a typical scattering mean free path is $\lambda_{\nu N}=(\sigma_{\nu N} n_0)^{-1}\simeq 25~{\rm cm}\,(100~{\rm MeV}/\epsilon_\nu)^2$. Energy is primarily carried by the species with the largest mean free path, in this case the heavy-lepton neutrinos, whereas the electron lepton number flux is carried, of course, by $\nu_e$ and $\overline\nu_e$.

\begin{figure}[b!]
\vskip-3pt
    \centering
    \includegraphics[width=0.90\columnwidth]{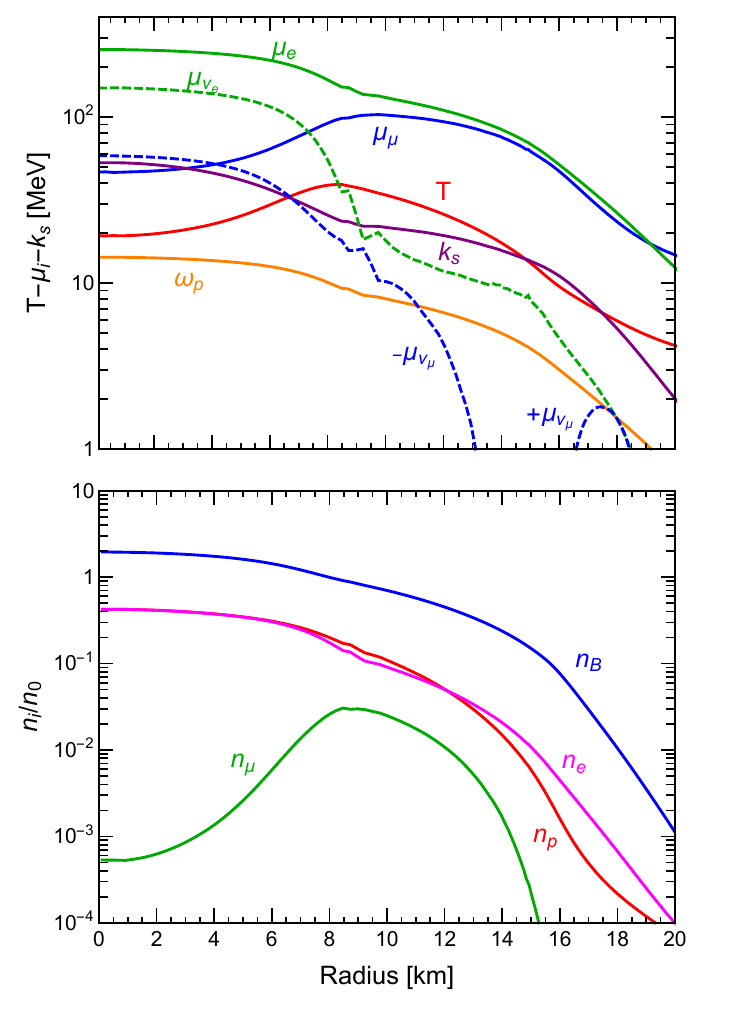}
    \caption{Profile of the Garching muonic SN model SFHo-18.8 at $t_{\rm pb}=1\,{\rm s}$ \cite{Bollig:2020xdr, Caputo:2021rux}. {\bf Top.} Chemical potentials, temperature, plasma frequency $\omega_{\rm p}$, and Debye screening scale $k_{\rm s}$. {\bf Bottom.} Number densities $n_i$, normalized to $n_0 = 0.181~{\rm fm}^{-3}$, corresponding to nuclear density of $3\times10^{14}~{\rm g}~{\rm cm}^{-3}$. (Figure from Ref.~\cite{Caputo:2021rux}, \copyright\ 2022 American Physical Society.)
    }
    \label{fig:profile}
\end{figure}

If neutrinos were locally in thermal and chemical equilibrium, the phase-space distribution (occupation number) for flavor $\alpha$ would be isotropic with a Fermi-Dirac distribution \smash{$f_{p,\alpha}=(e^{(\epsilon_{\nu_\alpha}-\mu_{\nu_\alpha})/T}+1)^{-1}$} with temperature $T$ and chemical potential $\mu_{\nu_\alpha}$ ($-\mu_{\nu_\alpha}$ for antineutrinos). All fermions follow analogous distributions and the condition of $\beta$ equilibrium implies $\mu_n-\mu_p=\mu_\alpha-\mu_{\nu_\alpha}$ for every flavor $\alpha=e$, $\mu$, and $\tau$. Typical values for these parameters can be gleaned from Fig.~\ref{fig:profile} that shows radial profiles for different key parameters for a numerical SN model at 1\,s pb. 

The crucial role of neutrinos in transporting energy and lepton number, on the other hand, derives from deviations from equilibrium, and moreover, flavor conversions may be crucial (Sec.~\ref{sec:FlavorEvolution}). In this more general context, one represents the distribution in phase space and in flavor space by $3\times3$ matrices $\varrho(\bp,\br,t)$; for antineutrinos $\overline\varrho(\bp,\br,t)$. They consist of the usual occupation numbers $(f_{\bp,\nu_e},f_{\bp,\nu_\mu},f_{\bp,\nu_\tau})$ for each flavor on the diagonal, and on the off-diagonal of complex-valued entries such as $\psi_{\nu_e,\nu_\mu}$ that encode flavor coherence caused by flavor oscillations. In the mean-field approach, the evolution of these matrices follows a well-known kinetic equation \cite{Dolgov:1980cq, Rudsky, Sigl:1993ctk, Fiorillo:2024fnl, Fiorillo:2024wej}
\begin{equation}\label{eq:QKE}
    (\partial_t+\bv\cdot\nabla_\br)\,\varrho_\bp=
    -i[{\sf H}_\bp,\varrho_\bp]+{\sf C}_\bp(\varrho,\overline\varrho),
\end{equation}
where neutrinos are assumed to be ultrarelativistic and have velocity $\bv=\bp/|\bp|$. An analogous equation pertains to anti\-neutrinos. The $3\times3$ matrices in flavor space $\varrho_\bp$, ${\sf H}_\bp$, and ${\sf C}_\bp$ depend on space and time, and the collision term ${\sf C}_\bp(\varrho,\overline\varrho)$ is very symbolically written---it depends on all $\varrho_\bp$ and $\overline\varrho_\bp$ at space-time point $(\br,t)$. It encodes the gain minus loss of neutrinos with a certain $\bp$ due to absorption, scattering, or pair-annihilation or production. References to the corresponding interaction rates are found in the appendix of Ref.~\cite{Buras:2005rp}, for the extension to muon neutrinos see Refs.~\cite{Bollig2018, Guo:2020tgx}, and for a numerical open-source library Ref.~\cite{OConnor:2015}. The Hamiltonian matrix ${\sf H}_\bp$ engenders coherent flavor evolution to be discussed later (Sec.~\ref{sec:FlavorEvolution}).

A significant practical limitation in simulating SN evolution lies in solving this transport equation. Deep inside a SN core, the neutrino mean free path is small compared to the characteristic spatial variation of the medium, and the distribution is essentially isotropic, with only a small perturbation that can be treated using the diffusion approximation of radiative transfer. Farther out, neutrinos stream freely. The intermediate regime---where neutrinos begin to decouple---is precisely the region that matters for the explosion, and it is geometrically extensive. In the Sun, where energy is transported by photons, radiation decouples in the photospheric region, which has a thickness of several 100~km, much smaller than the solar radius of $6.96\times10^5$~km, meaning that the Sun actually has a fairly well-defined surface. In a SN core, the neutrino decoupling region depends strongly on energy and flavor and extends over tens of km, comparable to the size of the PNS itself. The often-invoked picture of a neutrino sphere is a fairly rough concept. For $\nu_e$ and $\overline\nu_e$, it is formally defined as the radius where, for a given energy, the optical depth against $\beta$ processes is 2/3. For heavy-flavor states, one needs to distinguish between the transport sphere, where they last scatter, and the energy sphere, where they last undergo a pair process such as bremsstrahlung \cite{Raffelt:2001kv, Keil:2002in}.

Since muons play a secondary role for SN physics, SN neutrino transport is often simplified by using only three species, consisting of $\nu_e$, $\overline\nu_e$, and $\nu_x$, to accelerate the numerical solution. In this case $\nu_x$ represents any of $\nu_\mu$, $\overline\nu_\mu$, $\nu_\tau$, and $\overline\nu_\tau$, which are all forced to have the same phase-space distribution and whose interactions, being very similar, are treated identically. However, even ignoring muons, the heavy-lepton neutrinos and antineutrinos do not interact exactly identically, notably the elastic scattering rate on nucleons differs between, say, $\nu_\tau$ and $\overline\nu_\tau$ due to an effect called weak magnetism~\cite{Horowitz:2001xf}. In some SN simulations, a four-species transport is used that distinguishes between $\nu_x$ and $\overline\nu_x$. It is only recently that including muons and full six-species transport is becoming a new standard~\cite{Bollig:2017lki, Fischer:2020vie}, but these processes are not universally included, especially when computational power remains a limitation, as in three-dimensional simulations. The actual algorithm for neutrino transport includes further approximations to reduce the large dimensionality of the problem (three dimensions for $\bp$, three for $\br$, and time). The technical implementations differ between different codes---we refer to Refs.~\cite{Richers:2017awc, Glas:2018oyz, OConnor:2018sti, Mezzacappa:2020oyq} for discussions of the challenges and comparison of different methods.

Ordinary matter does not contain $\mu$ or $\tau$ leptons, but the pre-collapse iron core is electron rich. The nuclear charge $Z=26$ and typical mass number $A=56$ implies an electron fraction per baryon of $Y_e\simeq 26/56=0.46$. While electrons can convert to $\nu_e$ during collapse and partly escape, neutrino trapping prevents a substantial reduction of $Y_e$ so that the collapsed core maintains $Y_e\simeq 0.25$--$0.4$ and an electron lepton fraction (including $\nu_e$ and $\overline\nu_e$) of $Y_\mathrm{lep}\simeq 0.3$--$0.4$, depending on radial position. Subsequently, the SN core deleptonizes by diffusion and convection, i.e., this trapped electron lepton number is carried away by an excess $\nu_e$ flux within a few seconds. The large post-collapse $Y_e$ implies a large electron chemical potential of $\mu_e=150$--$200~{\rm MeV}$. 

The pronounced $\nu_e$--$\overline\nu_e$ asymmetry is another reason why the neutrino electron flavor plays a different role than heavy-lepton neutrinos. Including muons and six-species transport, a small amount of muon lepton number builds up by weak reactions producing muons and corresponding loss of $\overline\nu_\mu$ by diffusion. As an example for the inner PNS conditions, we show in Fig.~\ref{fig:profile} the profile of a muonic SN model at 1~s pb. The temperature maximum is still off center---it moves inward as deleptonization proceeds. Directly after bounce, the central regions are quite cool with a temperature of only a few MeV, whereas the hottest regions can reach $T=40$--$60~{\rm MeV}$, depending on the exact model and the nuclear equation of state.

\subsection{Expected neutrino signal}

\noindent The neutrino signal expected from a typical core-collapse SN shows a number of distinct phases. Even during the final phases of nuclear burning of the progenitor star, most of the liberated nuclear energy goes into neutrinos. During core silicon burning, a few days before collapse,  $L_\nu\simeq10^{46}~{\rm erg}~{\rm s}^{-1}$ and during Si shell burning, hours before collapse, it could be as high as $L_\nu\simeq10^{48}~{\rm erg}~{\rm s}^{-1}$ of sub-MeV neutrinos, mostly $\nu_e$ and $\overline\nu_e$ \cite{Odrzywolek:2010zz}. Still, for the red supergiants nearest to Earth, a pre-supernova signal could be detected and provide an early warning of an impending stellar collapse within the next few hours \cite{Odrzywolek:2010zz, Kato:2020hlc}.

\begin{figure*}[t]
    \centering
    \includegraphics[width=0.8\textwidth]{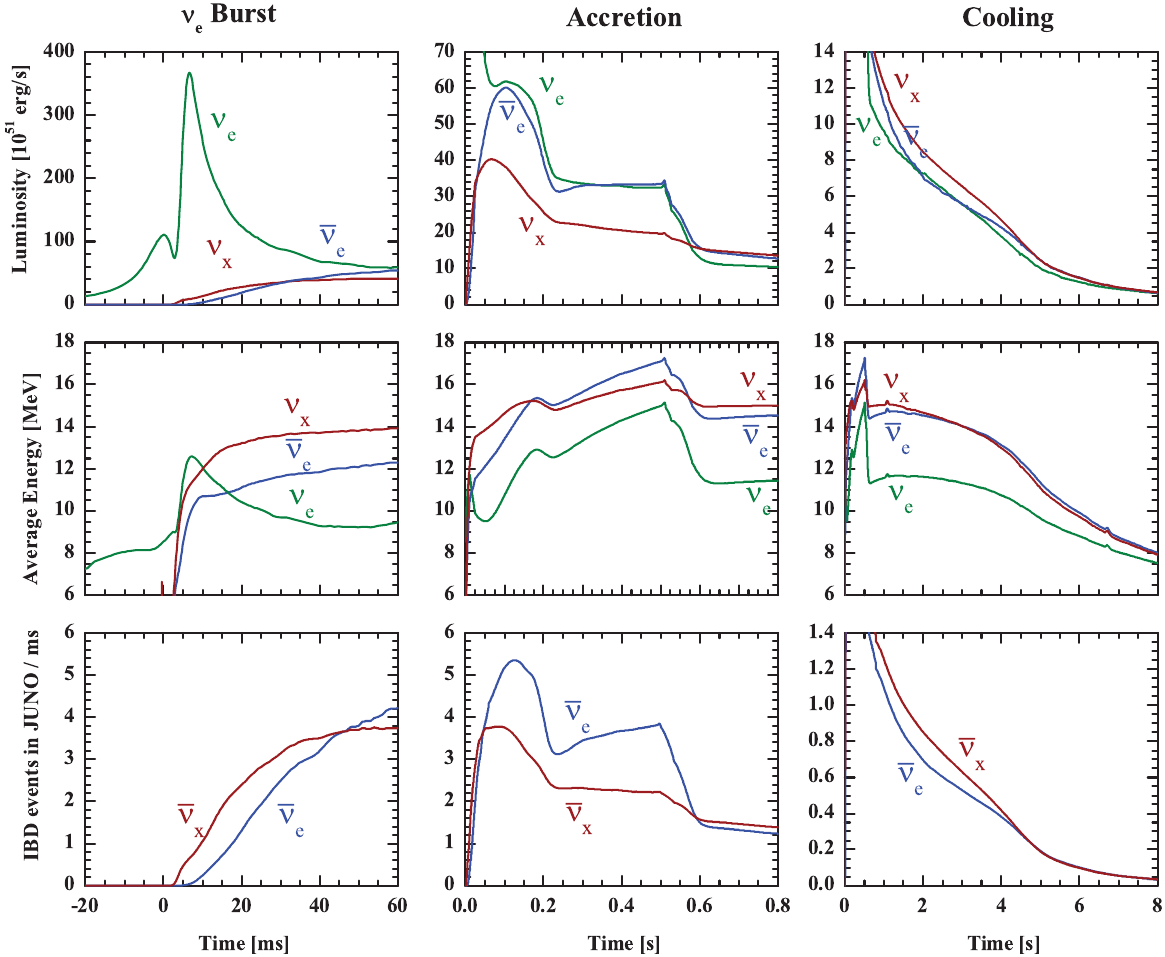}
    \caption{Neutrino signal of a typical core-collapse supernova (SN), based on a spherically symmetric Garching model with explosion triggered by hand during 0.5--0.6~ms \cite{Mirizzi:2015eza}. Shown are the flavor-dependent luminosities and average energies as well as the inverse beta rate ($\overline\nu_ep\to ne^+$) in the 20~kt JUNO scintillator detector (commissioned in 2025) for a SN distance of 10~kpc. No flavor conversion is assumed for the curves labeled $\overline\nu_e$, or complete flavor swap (curves $\overline\nu_x$). There are three main phases of neutrino emission, from left to right: (1)~Infall, bounce and initial shock propagation, including prompt $\nu_e$ burst. (2)~Accretion phase with significant flavor differences of fluxes and spectra and possible time variations of the signal. (3)~Cooling of the newly formed neutron star, only small flavor differences between fluxes and spectra.  (Figure adapted from Ref.~\cite{JUNO:2015zny} with permission, 
    \copyright\ 2016 IOP Publishing Ltd.)    
    }
    \label{fig:NeutrinoSignal}
\end{figure*}

When the core implodes, $L_\nu$ shoots up, and when the shock wave passes the outer layers of the original core, it dissociates several tenths of a solar mass of iron, absorbing enough energy to prevent a prompt explosion. At this stage, electron capture on protons ($e^-p\to n\nu_e$) and on remaining heavy nuclei releases a huge $\nu_e$ burst, called prompt neutrino burst or deleptonization burst or breakout pulse, shown in the left panels of Fig.~\ref{fig:NeutrinoSignal}. The peak luminosity of nearly $4\times 10^{53}~{\rm erg}~{\rm s}^{-1}$ and time profile of this burst are very characteristic and largely independent of input parameters \cite{Choi:2025igp, Kachelriess:2004ds, Wallace:2015xma}. Likewise, the rise of the other species is very characteristic and quantitatively different between $\overline\nu_e$ and $\overline\nu_x=\overline\nu_\mu$ or $\overline\nu_\tau$, the former being initially suppressed by the same large chemical potential in the source region that is responsible for the strong $\nu_e$ signal. Large SN detectors are usually most sensitive to IBD ($\overline\nu_e p\to ne^+$) so that a high-statistics observation of the $\nu_e$ burst is difficult. Therefore, the liquid argon detectors in the upcoming DUNE experiment (Sec.~\ref{sec:nuobservatories}) are of particular interest because they measure $\nu_e$ through the reaction $\nu_e+{}^{40}{\rm Ar}\to{}^{40}{\rm K}^*+e^-$. However, the entire $\nu_e$ burst could arrive almost entirely in the form of $\nu_x$, depending on the exact flavor oscillation scenario, reducing the $\nu_e$ signal to around 2\% of its original strength (Sec.~\ref{sec:FlavorEvolution}), preventing one from measuring the detailed $\nu_e$ burst profile. Also in IBD detectors, the characteristic onset behavior for $\overline\nu_e$ vs.\ $\overline\nu_x$ may allow one to distinguish between different flavor conversion scenarios \cite{Abbasi:2011, Serpico:2011ir}. In the bottom left panel of Fig.~\ref{fig:NeutrinoSignal}, we see the expected signal in JUNO, a 20~kt scintillator detector currently (2025) being commissioned, if there were no flavor conversion at all (curve $\overline\nu_e$) or complete flavor conversion (curve $\overline\nu_x$).

After the first phase of propagation, the shock stalls at a radius of some 150~km, while matter keeps falling through the shock, and this mass accretion powers neutrino emission, strongly favoring $\nu_e$ and $\overline\nu_e$ over the heavy-flavor species as we can see in the middle-column panels of Fig.~\ref{fig:NeutrinoSignal}. This accretion phase typically lasts for a few 100~ms until the Si/O interface of the progenitor star passes the shock-wave radius, followed by reduced accretion determined by the progenitor profile. At this point, the shock begins moving and starts the explosion. In the spherically symmetric simulation shown in Fig.~\ref{fig:NeutrinoSignal}, this effect was triggered by hand, causing an artificial explosion. The duration of the accretion phase and its mass-infall rate largely depend on the profile of the progenitor star, varying from case to case, depending on stellar mass, mass loss, rotation, or binary history. 

When accretion ends, the PNS cools by neutrino emission over several seconds. The luminosities in all six species converge, although the average $\nu_e$ energies are smaller caused by their larger opacity. When the luminosities are roughly equal ($L_{\nu_e}\simeq L_{\overline\nu_e}$) and the mean energies are smaller ($\langle\epsilon_{\nu_e}\rangle < \langle\epsilon_{\overline\nu_e}\rangle$),  the number flux of $\nu_e$ must be larger, in agreement with continued deleptonization. The duration and time profile of the emission is affected by detailed properties of the nuclear equation of state, neutrino opacities, and the effect of PNS convection. However, the order of magnitude can be understood from the diffusion distance for $n$ steps of a random walk, which is roughly $\sqrt{n}\,\lambda$, assuming a step size of the order of a typical neutrino scattering mean free path of $\lambda\simeq 25$~cm that was estimated after Eq.~\eqref{eq:sigma_NC}. To diffuse over a distance of 10~km, the approximate PNS radius, thus requires $n\simeq(10~{\rm km}/25~{\rm cm})^2\simeq10^9$ steps. With the speed of light of $3\times 10^{10}~{\rm cm}~{\rm s}^{-1}$, one step of 25~cm takes about $10^{-9}~{\rm s}$, so that it takes about 1~s to diffuse over 10~km. This estimate also reveals that neutrinos trapped in the PNS scatter with a typical rate~of~$10^9~{\rm s}^{-1}$.

\subsection{Strong deviations from spherical symmetry}
\label{sec:non-spherical}

\begin{figure*}[t]
\vskip15pt
    \centering
    \includegraphics[width=1.0\textwidth]{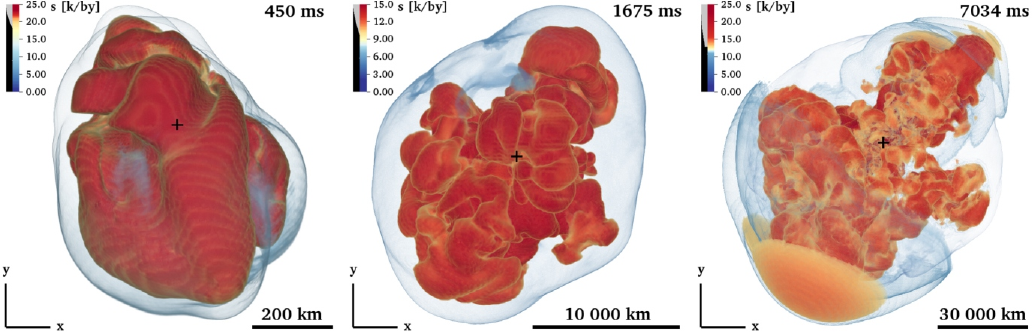}
    \caption{Explosion geometry of the 3D simulation of the Garching group \cite{Bollig:2020phc} shown in Fig.~\ref{fig:SN-3D}. What is shown here are 3D volume renderings of plumes of high-entropy neutrino-heated matter viewed from the $+z$-direction at the indicated times after core bounce. The surrounding, transparent, bluish surface is the supernova shock. The black cross marks the position of the proto-neutron star. At the last time shot ($7.034~{\rm s}$), the shock has entered the He layer and the entropy per baryon of postshock matter and neutrino-heated gas becomes similar. (Figure from \cite{Bollig:2020phc} with permission of the AAS.)}
    \label{fig:Explosion-3D}
\end{figure*}

\noindent While the simplest picture of core collapse and SN explosion is framed in spherical symmetry (1D), the explosions themselves are generically multi-dimensional phenomena and only few progenitors (with oxygen-magnesium cores or very small iron cores) near the low-mass end of stars undergoing core collapse are found to explode also in state-of-the-art 1D simulations \cite{Kitaura+2006, Janka:2006fh, Janka:2012wk, Muller:2016izw}.
The fact that there are actually large deviations from sphericity, i.e., pronounced three-dimensional (3D) effects, was recognized by the detailed observations of SN~1987A (Sec.~\ref{sec:SN1987A}), the closest SN in centuries, whose ejecta revealed clumpiness and large-scale asymmetries \cite{Arnett+1989}. Moreover, many NSs are observed to have large velocities (``natal kicks'') of several $100~{\rm km}~{\rm s}^{-1}$ and even up to $1000$--$2000~{\rm km}~{\rm s}^{-1}$, that are probably caused by the explosion asymmetry and subsequent gravitational interaction with the ejected matter (for recent comprehensive discussions, see Refs.~\cite{Burrows+2024, Janka+2024}). 

Over the past decades, numerical simulations have progressively explored multi-D effects both in 2D (axial symmetry) and fully in 3D (no assumed symmetries). While such simulations remain at the frontier of what is numerically feasible and only begin to permit large parameter explorations, they are quickly becoming the state of the art. In Fig.~\ref{fig:SN-3D} we already showed the numerical results from an exploding 3D model. In Fig.~\ref{fig:Explosion-3D} we display the spatial structure at several time shots, revealing explosion plumes and pronounced 3D structure. A visual rendition of a 3D explosion of a simulation by the Princeton group can be seen in a \href{https://www.youtube.com/watch?v=i-Ly8aCoF7E}{YouTube movie}, and several movies of 3D explosions by the Garching group are found under \href{https://wwwmpa.mpa-garching.mpg.de/ccsnarchive/movies/}{https://wwwmpa.mpa-garching.mpg.de/ccsnarchive/movies/}. The progenitor models used in such simulations are in most cases spherically symmetric and the large-scale convective structures arise from hydrodynamical instabilities in the neutrino-heating layer behind the stalled SN shock. (The model shown in Figs.~\ref{fig:SN-3D} and \ref{fig:Explosion-3D} is a rare exception: the pre-collapse model included asymmetries from convective oxygen-shell burning. The corresponding progenitor perturbations were crucial for the explosions because they strengthened convective activity in the neutrino-heated postshock region~\cite{Bollig:2020phc}.) Broadly speaking, numerical 3D models explode more easily than 1D ones because the convective circulation of matter allows to absorb more neutrino energy in the gain region, helping to rejuvenate the shock wave. Convection is also active inside the newly formed, still hot NS, accelerating the transport of energy and lepton number to the PNS surface.

\begin{figure}[b!]
\vskip-4pt
    \centering
    \includegraphics[width=0.95\columnwidth]{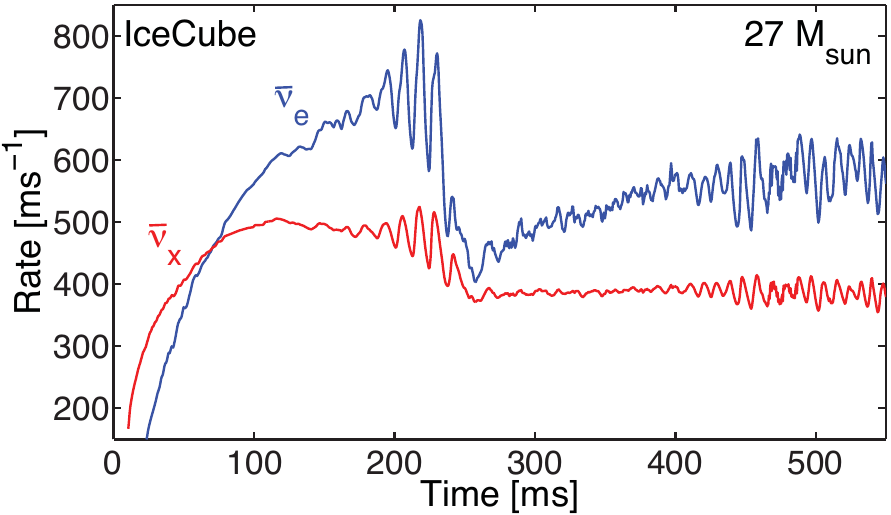}
    \caption{Inverse beta decay ($\overline\nu_e p\to n e^+$) detection rate in IceCube for a Garching numerical supernova model of a $27\,M_\odot$ star \cite{Tamborra:2013laa}, showing strong signal modulation by the SASI instability, i.e., the sloshing or spiral motion of the shock wave relative to the proto-neutron star imprints itself on the neutrino luminosity. The line labeled $\overline\nu_e$ is the signal in the absence of flavor conversion, whereas the line $\overline\nu_x$ obtains for complete conversion. The observer direction is chosen for an optimal effect---for a direction orthogonal to the sloshing or spiral motion, the effect would be much smaller.
    (Figure adapted from Ref.~\cite{Tamborra:2013laa},
    \copyright~2013 American Physical Society.)}
    \label{fig:SASI}
\end{figure}

In addition to these 3D effects, two dramatic hydrodynamical instabilities were numerically discovered, the SASI \cite{Blondin:2002sm, Blondin+2007} and LESA \cite{Tamborra:2014aua} instabilities, which impact also the neutrino signal. The former, the {\em Standing Accretion Shock Instability}, develops in the region between PNS and stalled shock during the accretion phase while the shock wave lingers at a radius of between 100 and 150~km and consists of pronounced dipolar oscillations or spiral motions (typical periods $10$--$20~{\rm ms}$) of the shock wave relative to the PNS. The modulated accretion rate onto the PNS also leads to a modulation of neutrino emission that could be detected in a high-statistics signal from the next Galactic SN, notably in the IceCube neutrino telescope at the South Pole and the Hyper-Kamiokande detector that is currently under construction in Japan (\cite{Tamborra:2013laa,Tamborra+2014}; Sec.~\ref{sec:Next-Galactic}). Both measure primarily the IBD process ($\overline\nu_e p\to n e^+$) that would show a pronounced oscillatory rate as can be seen in Fig.~\ref{fig:SASI}. This modulation can be a large effect, but depends on the observer direction; the effect is much smaller orthogonal to the sloshing motion and perpendicular to the plane of the SASI spiral motion. Moreover, whether a strong SASI motion develops depends on the duration of shock stagnation and possible phases of shock retraction, and thus on the progenitor star.  Core-collapse events with BH formation are particularly interesting cases for strong SASI effects. The LESA ({\em Lepton-number Emission Self-sustained Asymmetry}) instability is a dipole asymmetry of the convective layer inside the PNS that is stable or only slowly changing its direction. It thus causes a long-lasting large angular variation of the $\nu_e$ and $\overline\nu_e$ emission, meaning that lepton number is primarily emitted in one hemisphere. The impact of the SASI and LESA instabilities on the neutrino signal was systematically studied~\cite{Tamborra:2013laa, Tamborra+2014, Walk+2018, Walk+2019, Walk+2020, Nagakura:2020qhb}, showing that the instantaneous event rate in a given detector can vary by more than 50\%, depending on observer direction, but the time-integrated signal by less than 20\%.

Another possible consequence of nonspherical explosion and large-scale convective overturn is that matter is still infalling toward the NS in some regions, whereas in other directions gas is already explosively ejected. Moreover, there could be some amount of fallback seconds to minutes after the onset of the explosion, producing some amount of neutrino emission during or even after PNS cooling \cite{Chevalier1995,Akaho:2023alv}.

Yet another result of non-spherical dynamics is gravitational-wave (GW) emission that requires a time-varying quadrupole moment of the matter distribution or neutrino emission \cite{Mueller+1997, Mueller+2012, Andersson2019, Kalogera:2021bya}. GW observatories are seeing ever more binary mergers \cite{KAGRA:2021vkt, Roulet:2024cvl, LVK-2025}, whereas GWs from core-collapse SNe \cite{Ott:2008wt, Kotake:2011yv, Mueller+2013, Radice:2018usf, Kotake+2017, Andresen+2017, Andresen+2019, Powell:2018isq, Torres-Forne+2019, Zha+2020, Kuroda+2022, Jakobus+2023, Vartanyan:2023sxm, Mezzacappa:2024zph, Choi+2024, Murphy:2025vyv} remain to be observed. In nonrotating progenitors, one expects an initial GW burst from convection behind the decelerating SN shock shortly after bounce, caused by a negative entropy gradient. This activity calms down after a few 10~ms, generically followed by a relatively quiescent phase (around 100~ms) of low GW emission. Subsequently, violent mass motions caused by hydrodynamic instabilities and PNS oscillations dominate. Long-lasting PNS convection, fallback mass, the asymmetry of expanding ejecta, and possible nuclear phase transitions are other~sources. In rotating models, further large-scale hydrodynamic asymmetries provide yet more sources \cite{Moenchmeyer+1991, Dimmelmeier+2002a, Dimmelmeier+2002b, Kotake+2003, Dimmelmeier+2007, Kotake+2009, Kuroda+2016, Takiwaki+2021}. 

GWs from mass-quadrupole asymmetries have typical frequencies of tens to few 1000~Hz with amplitudes $A_\mathrm{m}$ of typically a few cm in 3D, corresponding to GW strains on the order of $h_\mathrm{m} \sim 10^{-22}$ for a Galactic source at $D = 10$~kpc. It was recently demonstrated that energy deposition by new particles at the bottom of the PNS convection zone can stimulate PNS convection, causing nonstandard GW emission during the otherwise more quiescent phase, signifying new physics \cite{Ehring:2024mjx}. 

Also quadrupole moments of anisotropically emitted neutrinos produce GW signals \cite{Epstein1978,Burrows+1996,Mueller+1997,Mueller+2012} with amplitudes
\begin{equation}
A_\nu(t) = 1.65\times 10^4 \int_{-\infty}^{t-D/c}dt'\,\alpha_\nu(t')\,\frac{L_\nu(t')}{100\,\mathrm{B}}\,\,\,\mathrm{cm}\,,
\end{equation}
where $L_\nu(t)$ is the time-dependent total neutrino luminosity and $\alpha_\nu(t)$ encodes the emission asymmetry. Maximum amplitudes in 3D models exceed  1000~cm ($h_\nu \sim 10^{-19}$ for $D = 10$~kpc), but their low frequencies around 1--10~Hz are hard to measure with current GW interferometers. 
\subsection{Diversity of supernova explosions and compact transient neutrino sources}
\label{sec:diversity}

\noindent The core-collapse SN phenomenon, as laid out in the previous sections, represents our current understanding in the framework of the Bethe--Wilson neutrino-driven mechanism, which is broadly---but not universally \cite{Soker:2024llc}---accepted as a standard paradigm. However, even if it is largely correct, significant variations are unavoidable, partly deriving from the progenitor star itself: its initial mass, subsequent mass loss, possible binary interaction, rotation, and magnetic field strength all play a role. 

The populations of Galactic and extragalactic X-ray binaries hosting a BH \cite{Gilfanov+2022} as well as recent GW observations of astrophysical BHs with masses of a few to several 10~$M_\odot$ \cite{KAGRA:2021vkt, Roulet:2024cvl, LVK-2025} indicate that some stellar collapses do not result in explosions (failed SNe). Even these events will produce substantial neutrino output during transient PNS cooling before final BH formation, as the compact remnant contracts and becomes increasingly hot \cite{Horiuchi:2017qja,Kresse+2021}. Observations from the ``death watch'' of a million red supergiants with the Large Binocular Telescope have identified at least one candidate for BH formation instead of a SN explosion~\cite{Kochanek:2023aob}. 

The key parameter distinguishing progenitors that will or will not explode is actively debated and should depend on details of pre-SN evolution. Besides core compactness (the mass-to-radius ratio of a defined core mass), also a critical-luminosity-based two-parameter criterion, a ``maximum fractional ram pressure jump discriminant'', a pronounced density jump near the Si/Si-O interface, the central entropy at collapse, and a more complex combination of different progenitor properties have been proposed \cite{OConnor:2010moj, Ertl+2016, Sukhbold:2015wba, Wang+2022, Boccioli+2023, Burrows:2024pur, Maltsev+2025}. However, all of these criteria are challenged by the present non-convergent results of 2D and 3D core-collapse simulations of different groups~\cite{Janka:2025tvf}.

The final NS mass range is broad, typical values being 1.2--$1.8\,M_\odot$ \cite{Ozel:2016oaf}. The largest measured one is the black-widow pulsar PSR J0952-0607 with a gravitational mass of $2.35\pm0.17\,M_\odot$ \cite{Romani:2022jhd}, whereas the lightest one is $1.174\pm 0.004\,M_\odot$, the lighter companion of the compact binary radio pulsar system J0453+1559 \cite{Martinez:2015mya}.

Among actually exploding SNe, there is a large variation of astronomical appearance \cite{Gal-Yam2017}. Spectral classification distinguishes types Ia–c (lacking hydrogen lines) from type II (showing them). Subtype Ia displays strong silicon lines but no hydrogen or helium; Ib lacks hydrogen; and Ic lacks both hydrogen and helium. Physically, type Ia SNe are interpreted as thermonuclear explosions of WDs, whereas types Ib, Ic, and II result from core collapse of massive stars. About one third of all SNe are of type Ia \cite{Cappellaro:1996cc,Graur+2017a,Graur+2017b}; these are brighter, more uniform, and are widely used as standardizable candles for cosmological distance determination, but they lack strong neutrino emission \cite{Odrzywolek:2010je}. Furthermore, based on the evolution of their light curves, type II SNe are subdivided into II-P (“plateau”) and II-L (“linear”). II-P SNe exhibit a nearly constant luminosity for roughly 80–100~days after maximum light, caused by hydrogen recombination in the extended envelope of a red supergiant progenitor. II-L SNe, by contrast, show a steady, almost linear decline in brightness, reflecting a smaller or more stripped hydrogen envelope.

The typical energy release of 200--400~B (recall that $1~{\rm B} = 10^{51}~{\rm erg} = 10^{44}~{\rm J}$) is the difference between the baryonic and gravitational final NS mass, but the actual value depends both on the mass itself and the uncertain nuclear equation of state (or the NS's final radius). Most of this huge amount of energy is released in the form of neutrinos, whereas the actual explosion energy is typically 1~B, or at most a few B, whereas the electromagnetic output is yet smaller, around 0.01~B---unless the SN shock dissipates a significant fraction of its kinetic energy in the dense circumstellar medium (CSM), thus causing enhanced light emission in so-called interacting SNe. 

However, the class of low-energy SNe has explosion energies as small as 0.1~B or even less, probably correlated with small NS masses, for example in electron-capture SNe. These originate from the core collapse of progenitors with highly degenerate oxygen-neon cores instead of iron \cite{Kitaura+2006, Muller:2016izw, Stockinger+2020}. 

On the other extreme are hypernovae with up to 10--50~B, possibly associated with very massive, rapidly rotating stars and often linked to long-duration gamma-ray bursts (GRBs) \cite{Woosley+2006}. Their extreme energies as well as the possible formation of ultrarelativistic GRB-jets could be explained by fast rotation of the collapsing stellar cores. A large magnetic field---as in a proto-magnetar or in a BH-torus system---can play an important role in translating rotation energy into explosion energy by a magnetorotational mechanism. Such extreme events surely involve physics beyond the Bethe--Wilson mechanism. Interaction with dense CSM or spin-down energy release of a proto-magnetar are also invoked to explain the very rare class of superluminous SNe (SLSNe) \cite{Gal-Yam2019, Moriya2024}.

As mentioned earlier, some of the lowest-energy explosions may correspond to the class of electron-capture SNe at the low end of the mass range for core collapse. These progenitors never form iron cores, but during the preceding evolution develop such large electron degeneracy in their O--Ne--Mg cores that electron captures on Mg, Na, and Ne and heavier nuclei in nuclear statistical equilibrium absorb enough electrons to trigger the collapse. While the existence of such events strongly depends on metallicity and binarity \cite{Podsiadlowski+2004, Poelarends+2008, Langer2012}, they can contribute to the population of low-energy SNe, but they are not necessarily faint because of an extremely inflated, low-density hydrogen envelope around the degenerate progenitor core or a dense progenitor wind~\cite{Tominaga+2013, Moriya+2014}.

Another hypothetical class of core-collapse explosions is connected to accretion-induced or merger-induced collapse (collectively termed AICs), which occurs when a WD reaches its stable mass limit, either by accretion from a binary companion or as remnant of a binary-WD merger \cite{Nomoto1986, Nomoto+1991, Woosley+1992, Fryer+1999, Dessart+2006}. Rather than being disrupted in a thermonuclear SN, the compact object entirely collapses to a NS, producing a faint transient with an ejecta mass of 0.005--0.05\,$M_\odot$ and a low explosion energy---similar to electron-capture SNe. AIC events may contribute to the population of low-kick NSs and have been suggested as sources of GRBs and kilonovae, but there is no definite evidence that they occur with significant rates (for recent works, see \cite{ChiKitCheong+2024, Batziou+2025}).

Yet another hypothetical class of SN explosions may occur at the opposite end of the mass spectrum, roughly between 150 and $250\,M_\odot$. In these stars, the cores before oxygen ignition become so hot that electron-positron pairs are efficiently thermally produced, reducing pressure support and triggering partial collapse followed by a violent explosion---this is known as a pair-instability SN, which leaves no compact remnant. Such events are especially plausible for the metal-poor early generations of stars \cite{Heger+2003,Langer2012}.

Detectable MeV-range months-long neutrino signals could arise from common-envelope systems \cite{Esteban:2023uvh}, where a NS is engulfed by the envelope of a binary companion, although the existence and frequency of such systems is unknown. Still, even the existing record of neutrino data should be analyzed for such signatures. They are similar to the hypothetical Thorne--\.{Z}ytkow objects \cite{1975ApJ...199L..19T, OGrady:2024ioe}, corresponding to a NS immersed in a supergiant envelope. Eventually, these hypothetical objects might also be detected in neutrinos, with a reach possibly as far as the Small Magellanic Cloud \cite{Martinez-Mirave:2025pnz}.

Neutrino emission comparable to that of a core-collapse SN obtains from a binary NS merger, where an accretion torus and hypermassive NS forms that may collapse to a BH after fractions of a second to seconds, depending on the mass of the merging binary, the nuclear equation of state and the angular momentum and its transport in the merger remnant. A prototype is the celebrated event of GW170817 that was detected in GWs on 17 August 2017 in the galaxy NGC 4993 at a distance of around 40~Mpc, and was also observed as the gamma-ray burst GRB170817A and the kilonova AT2017gfo \cite{LIGOScientific:2017ync,Abbott+2017}. Such mergers were long expected to produce GWs and GRBs as well as kilonovae with electromagnetic energies on the order of $10^{-4}$\,B, much smaller than the bolometric emission from SNe, but with comparable kinetic energies of the ejecta, exceeding 1\,B in extreme cases \cite{Metzger+2010, Goriely+2011, Bauswein+2013, Grossman+2014, Metzger+2014, Kasen+2017}. In a galaxy like the Milky Way, one expects between one and 25 such events per million years~\cite{GWTC-4.0:2025}, to be compared with a core-collapse SN rate of about two per century (Sec.~\ref{sec:SNrate}). Therefore, the future observation of a binary-merger neutrino signal is rather unlikely.

\subsection{Nucleosynthesis}

\noindent Practically all elements heavier than helium were produced in stars \cite{Burbidge:1957vc, Cameron:1957}, whereas all of the hydrogen, deuterium, some lithium, and most of the helium were produced in the early universe. The binding energy per nucleon achieves its maximum for iron-group elements, with 8.795~MeV being the absolute maximum for $^{62}$Ni and a very similar 8.791~MeV for $^{56}$Fe. Beyond these elements, nuclear fusion (``burning'') is endothermic, not exothermic, setting an inevitable endpoint to hydrostatic stellar evolution and making stellar collapse unavoidable as discussed~earlier.

On the other hand, the nuclear binding energy of $^4$He with 7.07~MeV per nucleon is so large that this nucleus can be seen as a particle onto itself, the $\alpha$ particle. Nuclear fusion in stars largely consists of assembling more and more $\alpha$ particles to higher-mass ``alpha conjugate'' nuclei such as $^{12}$C, $^{16}$O, and $^{20}$Ne, whereas no stable isotope with mass number 8 exists. On the other hand, large stable nuclei are neutron rich, so the most massive alpha conjugate stable nucleus is $^{40}$Ca, whereas the next one, $^{44}$Ti, is unstable to electron capture with a half life of 59~years. 

Therefore elements beyond the iron group can emerge neither from hydrostatic nuclear burning during ordinary stellar evolution nor from so-called explosive nuclear burning at the more extreme conditions provided by thermonuclear and core-collapse SNe or less spectacular transients such as novae (accreting WDs) and X-ray bursters (accreting NSs). In addition, it is not enough to just produce the elements, they also need to be dispersed in space, which is not the case, for instance, for the precollapse iron core: the iron is dissociated in the collapse, the necessary energy coming from gravity. While the iron-group elements belong to the most abundant, they must derive from explosive nucleosynthesis in exploding WDs and massive stars.

The direct result of nucleosynthesis in SNe is observed in their light curves. For example in the case of SN~1987A, after the initial maximum connected to hydrogen recombination, the bolometric luminosity is driven by the decay $^{56}{\rm Ni}\to{}^{56}{\rm Co}$ (half life 6.1~d) and then $^{56}{\rm Co}\to{}^{56}{\rm Fe}$ (half life 77.2~d) for the first few hundred days. During this period, it follows precisely the latter exponential decay law and corresponds to a produced $^{56}{\rm Ni}$ mass of $0.071\,M_\odot$ \cite{Seitenzahl:2014zia}, whereas later longer-lived isotopes such as $^{44}$Ti dominate. (Both $^{44}$Ti and $^{56}$Ni are unstable alpha-conjugate~nuclei.) 

The main message is that numerical models of SN~1987A not only need to produce the explosion, but also the observed amounts of radioactive nuclei~\cite{Sieverding:2023lju}. Another long-lived isotope produced in massive stars and their SNe is $^{26}$Al (half life $7.2\times10^5$~yr) that emits a characteristic 1.81~MeV gamma-ray in its decay sequence~\cite{Diehl+2021}. It has been measured by gamma-ray satellites, leading to an estimate of the Galactic SN rate over the past million years~\cite{Diehl:2006cf}---see Sec.~\ref{sec:SNrate}. Yet another interesting isotope is $^{60}$Fe (half life 2.6 million years) that has been detected in terrestrial deep-sea ferromanganese crust and lunar material \cite{Knie+2004,Fimiani+2016}, and possibly signifies a close-by SN over such a period~\cite{Breitschwerdt+2016}.

Nucleosynthesis of heavy elements in explosive astrophysical events can proceed via several distinct pathways. Explosive nuclear fusion reactions in sufficiently hot ejecta produce iron-group and intermediate-mass elements. Thus mainly $^{56}$Ni and silicon are made by carbon and oxygen burning behind the deflagration or detonation front in thermonuclear WD explosions. Similarly, the outgoing shock in core-collapse SNe reprocesses the carbon, oxygen, and neon layers of massive stars, creating some of the ejected $^{56}$Ni and nuclei mostly also assembled during hydrostatic burning~\cite{Woosley:2002zz, Sukhbold:2015wba}. 

Neutron-rich species in the trans-iron region can only be built up via neutron captures, because nuclear binding energy is released when free neutrons combine with preexisting heavy nuclei (``seed nuclei''). Thus elements along the valley of stability are generated by the slow neutron-capture process (s-process) in evolving, massive stars, where neutron captures are slower than beta decays. At neutron densities roughly 10 orders of magnitude higher, which occur in ejected PNS or NS matter, neutron captures happen more rapidly than beta decays (or, at high entropies, photodisintegration by energetic $\gamma$'s), and the heaviest of all elements (actinides and lanthanides including platinum, gold, thorium, and uranium) are created by the rapid neutron-capture process (r-process)~\cite{Cowan+2004}. 

In contrast, trans-iron nuclides on the proton-rich side of the valley of stability (p-nuclei) can form by proton-captures (\hbox{p-process}) in ejecta with proton excess, possibly fostered by neutron captures produced by $\bar\nu_e$ absorption on protons in the so-called $\nu$p-process~\cite{Froehlich+2006,Pruet+2006}. Rare proton-rich isotopes not formed by the other mechanisms could be made in the $\gamma$-process, which involves nuclear photodisintegration, where high-energy gamma rays knock nucleons and $\alpha$ particles out of heavy nuclei during shock heating of the oxygen and neon layers \cite{Arnould1976, Woosley+1978, Roberti+2023, Roberti+2024}. 

The $\nu$-process arises uniquely from the intense neutrino flux in core-collapse SNe, where neutrino interactions (charged-current absorption and neutral-current induced spallation) produce certain, partly rare, isotopes of boron, fluorine, lanthanum, and tantalum in the O/Ne, O/C, and He shells of the exploding star~\cite{Woosley+1990,Sieverding+2018,Sieverding+2019}. 

Finally, the $\alpha$-process consists of the accretion of $\alpha$ particles by seed nuclei, which then grow to larger masses.

The efficiency or even occurrence of the different pathways depends on the environment. The $\alpha$-process plays a particularly important role in any r-process that starts at high temperatures ($T\gtrsim 5$\,GK), with free nucleons in nuclear statistical equilibrium assembling to $\alpha$ particles when the ejected matter expands and cools. These $\alpha$ particles can further combine to heavier elements up to the iron group, thus forming the seed nuclei for subsequent neutron or proton capture in a primary r-process or $\nu$p-process. 

For a long time, the neutrino-driven wind---a high-entropy baryonic mass outflow driven off the PNS surface through energy deposition by the emitted neutrinos~\cite{Duncan+1986}---was thought to be a good site for the r-process \cite{Meyer+1992,Woosley+1994,Hoffman+1997}. However, the necessary high entropies of well above 100\,$k_\mathrm{B}$ per nucleon are hard to reach \cite{Witti+1994,Arcones+2007,Roberts+2010} and state-of-the-art SN models yield an excess of protons in the wind instead of neutron-rich conditions~\cite{Mirizzi:2015eza}. The absorption of $\nu_e$ and $\bar\nu_e$ determines the electron faction $Y_e^\mathrm{wind}$ in the neutrino-driven wind~\cite{Qian+1996},
\begin{equation}
 Y_e^\mathrm{wind} \approx \left(1 + \frac{L_{\bar\nu_e}}{L_{\nu_e}}\,
 \frac{E_{\bar\nu_e}-2\Delta + 1.2\Delta^2/E_{\bar\nu_e}}{E_{\nu_e}+2\Delta + 1.2\Delta^2/E_{\nu_e}} \right)^{\!\!-1} ,
 \label{eq:yewind}
\end{equation}
where $L_{\nu_i}$ are the neutrino luminosities, $E_{\nu_i} = \langle\epsilon_{\nu_i}^2\rangle/\langle\epsilon_{\nu_i}\rangle$ neutrino energies computed from the mean energies $\langle\epsilon_{\nu_i}\rangle$ and the mean squared energies $\langle\epsilon_{\nu_i}^2\rangle$ of the radiated neutrino spectra, and $\Delta = 1.293$\,MeV is the neutron-proton mass difference. The r-process in the wind is not only hindered by terms of order the neutrino energy over the nucleon mass neglected in Eq.~(\ref{eq:yewind})~\cite{Horowitz+1999}, but also by PNS convection, which boosts the $\nu_e$ luminosity compared to that of $\bar\nu_e$ until late times when the wind has become very weak~\cite{Mirizzi:2015eza}. 

Moreover, modern 3D simulations show that neutrino-driven winds are not a general phenomenon in core-collapse SNe, but may be constrained to explosions of the lowest-mass progenitors~\cite{Janka+2022}. In more massive stars, matter inflows to the PNS exist for many seconds, get heated by neutrinos and re-ejected, thus contributing to a continous growth of the explosion energy~\cite{Bollig:2020phc,Janka:2025tvf}. The nucleosynthesis conditions in such turbulent outflows are dramatically different from those in steady, spherical neutrino-driven winds due to much lower entropies and a non-monotonic temperature evolution~\cite{Sieverding:2023lju}. Despite a rather wide distribution of $Y_e$ in the ejecta---some slightly neutron-rich matter is admixed to dominantly proton-rich material---mostly the production of iron-group nuclei, in particular the major yield of $^{56}$Ni, is expected, and trans-iron nuclides only up to nuclear mass numbers around 100~\cite{Wang+2024}.

Today, NS mergers are thought to be the main source of r-process material including lanthanides and actinides. This has been predicted for a long time~\cite{Lattimer+1976,Symbalisty+1982,Meyer+1989,Freiburghaus+1999,Goriely+2011a,Goriely+2011} and may be observationally confirmed now, at least for the production of lanthanides, by the kilonova seen in connection with the NS merger event of GW170817 \cite{Kasen+2017,Smartt+2017,Watson+2019}. However, specific patterns in the chemical enrichment history of the Milky Way, deduced from metal-poor stars, are interpreted as signatures of other, additional sources of r-process matter \cite{Cowan+2021,Farouqi+2025}. These could be jets in magnetorotational SN explosions \cite{Nishimura+2006,Nishimura+2015,Moesta+2018,Arcones+2023,Reichert+2023}, BH accretion disks in rapidly rotating, collapsing massive stars \cite{Siegel+2019,Siegel2022}, magnetar giant flares \cite{Patel+2025,Patel+2025a}, or AIC events of rapidly spinning WDs~\cite{Batziou+2025}.  

In most of these environments, $Y_e$ in the ejecta is governed by $\nu_e$ and $\bar\nu_e$ emission and absorption (for NS mergers and their remnants, this is discussed in \cite{Wanajo+2014, Just+2022}), and the ejecta conditions depend on the fluxes, spectra, and flavor content. Since neutrino interactions play such a crucial role in SN nucleosynthesis, flavor conversion or, for instance, conversion to sterile states can cause large modifications. Flavor evolution and the detailed properties of the neutrino fluxes shaping the nuclear composition of the expanding SN ejecta are topics that are undergoing a transformation at present (Sec.~\ref{sec:FlavorEvolution}). The specific impact of neutrino conversion---and possibly new particles---on explosive nucleosynthesis in SNe and other environments are intriguing topics beyond the scope of this review.

\section{Neutrino flavor evolution}
\label{sec:FlavorEvolution}

\noindent Neutrinos famously oscillate between different flavors due to their large mass mixings, as elaborated elsewhere in this {\em Encyclopedia}. One might expect this effect to strongly impact the neutrino-driven core-collapse SNe. However, the large refractive potential from the medium renders neutrino interaction states practically identical to propagation eigenstates; neutrinos propagate as if they had no mixing at all \cite{Wolfenstein:1979ni}. Flavor conversion is ignored in all contemporary SN simulations, so the flavor density matrices $\varrho_\bp$ in Eq.~\eqref{eq:QKE} are taken to be purely diagonal---neutrinos are described by their flavor-eigenstate occupation numbers $f_{\bp,\alpha}$. They would later be affected by mixing in the stellar envelope, encountering MSW resonances \cite{Dighe:1999bi}, which adiabatically turn them to mass eigenstates. When detected through IBD ($\overline\nu_ep\to n e^+$), they are projected back on $\overline\nu_e$. Propagation through Earth would lead to further oscillations, imprinting a detector- and energy-dependent modulation on the spectrum \cite{Dighe:1999bi, Lunardini:2001pb, Dighe:2003jg, Fogli:2001pm, Borriello:2012zc, Hajjar:2023knk}. In fact, nonadiabatic propagation effects could already imprint themselves during propagation through the SN matter, for example, when jumps of density are encountered caused by shocks or reverse shocks \cite{Schirato:2002tg, Takahashi:2002yj, Fogli:2003dw, Tomas:2004gr, Dasgupta:2005wn} or matter turbulence \cite{Fogli:2006xy, Friedland:2006ta, Kneller:2010sc, Borriello:2013tha, Lund:2013uta}.

On the level of the kinetic equation Eq.~\eqref{eq:QKE}, flavor conversion is engendered by the Hamiltonian matrix, written in the weak-interaction basis and for ultrarelativistic neutrinos, 
\begin{equation}\label{eq:Hmatrix}
    {\sf H}_\bp=\frac{{\sf M}^2}{2\epsilon_\bp}+\sqrt{2}\GF{\sf N}_\ell+\sqrt{2}\GF\int\frac{d^3\bp'}{(2\pi)^3}\varrho_{\bp'}\,(1-\bv\cdot\bv'),
\end{equation}
where the first term, proportional to the square of the neutrino mass matrix ${\sf M}$, drives vacuum oscillations and is what breaks flavor lepton number conservation; $\epsilon_\bp=|\bp|$ is the energy in the ultrarelativistic limit. For antineutrinos, this term appears with a minus sign. The second term, proportional to the diagonal matrix of charged-lepton number densities ${\sf N}_\ell$, is responsible for the neutrino energy shift by matter refraction. The third term represents neutrino-neutrino refraction. It shows explicitly the dependence on neutrino direction in that those with equal directions of motion ($\bv=\bv'$) do not affect each other; the influence varies as $(1-\cos\theta)$ with relative direction.

The three neutrino masses are separated by two mass differences, the solar one of $\Delta m^2_{\rm sol}=(8.6~{\rm meV})^2$, which drives the original solar neutrino conversion, and the atmospheric one of $\Delta m^2_{\rm atm}=(50~{\rm meV})^2$, responsible for the original flavor conversion of atmospheric neutrinos. Cancelation of the vacuum and matter terms in Eq.~\eqref{eq:Hmatrix} corresponds to an MSW resonance that requires $(\Delta m^2/2\epsilon_\nu)\cos2\theta=\sqrt2\GF n_e$, where $\theta$ is the two-flavor mixing angle and $n_e$ the electron density. Passing through the progenitor star, neutrinos thus encounter two such resonances. Assuming energy $\epsilon_\nu=10$~MeV and the solar neutrino mixing angle $\theta_{12}=33.7^\circ$ ($\cos2\theta_{12}=0.386$) imply that the L resonance occurs for $\rho_{\rm L}= 38~{\rm g}~{\rm cm}^{-3}\,(0.5/Y_e)\,(10~{\rm MeV}/\epsilon_\nu)$. From solar neutrino measurements it is known to occur among neutrinos (not antineutrinos), meaning normal mass ordering in the 1--2 sector. The mixing angle relevant for $\nu_e$--$\nu_x$ conversion is $\theta_{13}=8.5^\circ$ so that $\cos2\theta_{12}\simeq1$ and the H resonance occurs for $\rho_{\rm H}=3.2\times10^3~{\rm g}~{\rm cm}^{-3}\,(0.5/Y_e)\,(10~{\rm MeV}/\epsilon_\nu)$. However, it remains unknown if this resonance occurs in the $\nu$ or $\overline\nu$ sector, i.e., this mass ordering remains unknown and is one of the few pieces still missing from the neutrino mass and mixing matrix. 

In the absence of other effects, a distant detector sensitive to the $e$ flavor will measure a SN neutrino flux of \cite{Dighe:1999bi}
\begin{subequations}
\begin{eqnarray}
    F_{\nu_e}^{\rm det}&=&p\,F_{\nu_e}^0+(1-p)\,F_{\nu_x}^0,
    \\
    F^{\rm det}_{\overline\nu_e}&=&\overline p\,F_{\overline\nu_e}^0+(1-\overline p)\,F_{\overline\nu_x}^0,
\end{eqnarray}
\end{subequations}
where the superscript 0 refers to the flux emerging from the neutrino sphere, and $p$ or $\overline p$ are the $\nu_e$ and $\overline\nu_e$ survival probabilities. If the propagation is fully adiabatic, and for normal mass ordering (NO)---where both resonances occur in the $\nu$ sector---a $\nu_e$ produced at high density emerges as a $\nu_3$ mass eigenstate, whereas a $\overline\nu_e$ emerges as $\overline\nu_1$. Conversely, for inverted ordering (IO)---H~resonance in $\overline\nu$ sector---a produced $\nu_e$ emerges as $\nu_2$ and a $\overline\nu_e$ as $\overline\nu_3$ \cite{Dighe:1999bi, Vitagliano:2019yzm}. After projecting back on the $\nu_e$ or $\overline\nu_e$ state for a charged-current detector, the survival probabilities are given in Table~\ref{tab:survival}. In other words, one will never observe the originally emitted spectra. In particular, in the $\overline\nu$ sector, the early rise time of the SN burst could reveal the neutrino mass ordering, because one would see a different combination of original fluxes~\cite{Serpico:2011ir}. On the other hand, in NO, the prompt $\nu_e$ burst would almost entirely appear as $\nu_x$.

\begin{table}[ht]
\centering
	\caption{SN $\nu_e$ and $\overline\nu_e$ survival probability for adiabatic propagation.\label{tab:survival}}
    \begin{tabular}{@{}lll}
			\toprule
			&Normal Ordering&Inverted Ordering\\
			\midrule
			$p$&$\sin^2\theta_{13}\simeq0.022$&$\sin^2\theta_{12}\cos^2\theta_{13}\simeq0.30$ \\
			$\overline p$&$\cos^2\theta_{12}\cos^2\theta_{13}\simeq0.68$
			&$\sin^2\theta_{13}\simeq0.022$\\
			\bottomrule
	\end{tabular}
\end{table}

However, this simple picture is now largely outdated. Flavor evolution is highly nontrivial and cannot simply be post-processed on a given SN background model, primarily due to neutrino-neutrino refraction itself \cite{Pantaleone:1992eq}, which allows for collective flavor conversions \cite{Samuel:1993uw, Samuel:1995ri, Duan:2006an, Sawyer:2004ai, Sawyer:2008zs, Izaguirre:2016gsx}. The history of this subject has undergone multiple paradigm shifts, and is too convoluted to detail here; see \hbox{Refs.~\cite{Duan:2009cd, Duan:2010bg}} for early reviews and Refs.~\cite{Tamborra:2020cul, Richers:2022zug, Johns:2025mlm} for more recent ones. The key insight is that, at large densities, individual neutrino lepton numbers are conserved, but pairwise conversion is possible---for example, processes such as $\nu_e\overline\nu_e\to \nu_\mu\overline\nu_\mu$ or $\nu_e(\bp)\nu_\mu(\bp')\to \nu_e(\bp')\nu_\mu(\bp)$---meaning that flavor is redistributed in phase space without net conversion. Such pairwise conversion can happen via direct neutrino-neutrino collisions, yet there is a much more efficient channel, namely the coherent production of a weak field, encoded in the refractive term. 

This process is what defines a collisionless plasma; in a standard (electronic) collisionless plasma, electrons do not exchange energy by pairwise Coulomb scattering, but rather by producing a coherent electric field. Thus, neutrinos in SNe form a plasma state, and their coherent flavor dynamics can significantly impact the SN evolution, owing to the special role of the trapped electron lepton number and the strong energy dependence of neutrino interaction rates. The key challenge, preventing a quantitative description of the process, is that the coherent field thus produced changes on much smaller temporal and spatial scales than the hydrodynamic ones, forbidding a brute-force numerical solution.

The mean fields of neutrino flavor coherence, which in linear theory are the off-diagonal terms of the flavor density matrices in Eq.~\eqref{eq:QKE}, can be pictured as independent dynamical degrees of freedom. They have dispersion laws of their own, defined by the state of the neutrino plasma, and can develop unstable modes, whose amplitude grows exponentially. The analogy with an electronic plasma is again instructive; an excess of free energy, caused for example by a beam of energetic electrons in a background plasma, can be released by producing a turbulent electric field, which can also be seen as a collection of plasmons. 

Likewise, in a neutrino plasma dominated by one flavor---usually in SNe this would be $\nu_e$---the appearance of neutrinos with the opposite lepton number---which can either be $\overline{\nu}_e$ or $\nu_\mu$, which we might call flipped neutrinos---creates an excess of free energy that can be released by an unstable flavor field \cite{Fiorillo:2024bzm, Fiorillo:2024uki, Fiorillo:2024pns, Fiorillo:2025ank, Fiorillo:2025zio}. Flavor instabilities require then a ``crossing'' \cite{Dasgupta:2021gfs, Johns:2024bob, Fiorillo:2024bzm, Fiorillo:2024dik}, i.e., a change of sign as a function of energy and/or direction of motion of the difference between, say, electron and muon lepton number carried by neutrinos, triggered by the flipped ones. A crossing in the energy-integrated angular distribution is sufficient to cause an instability even for massless neutrinos \cite{Morinaga:2021vmc, Dasgupta:2021gfs, Fiorillo:2024bzm, Fiorillo:2024dik} (fast instability), while neutrino masses can render unstable even a distribution with a crossing in the energy distribution along any direction \cite{Fiorillo:2024pns, Fiorillo:2025ank, Fiorillo:2025zio, Fiorillo:2025kko} (slow~instability).

\begin{figure}[b]
	\centering
   \includegraphics[height=3.0cm]{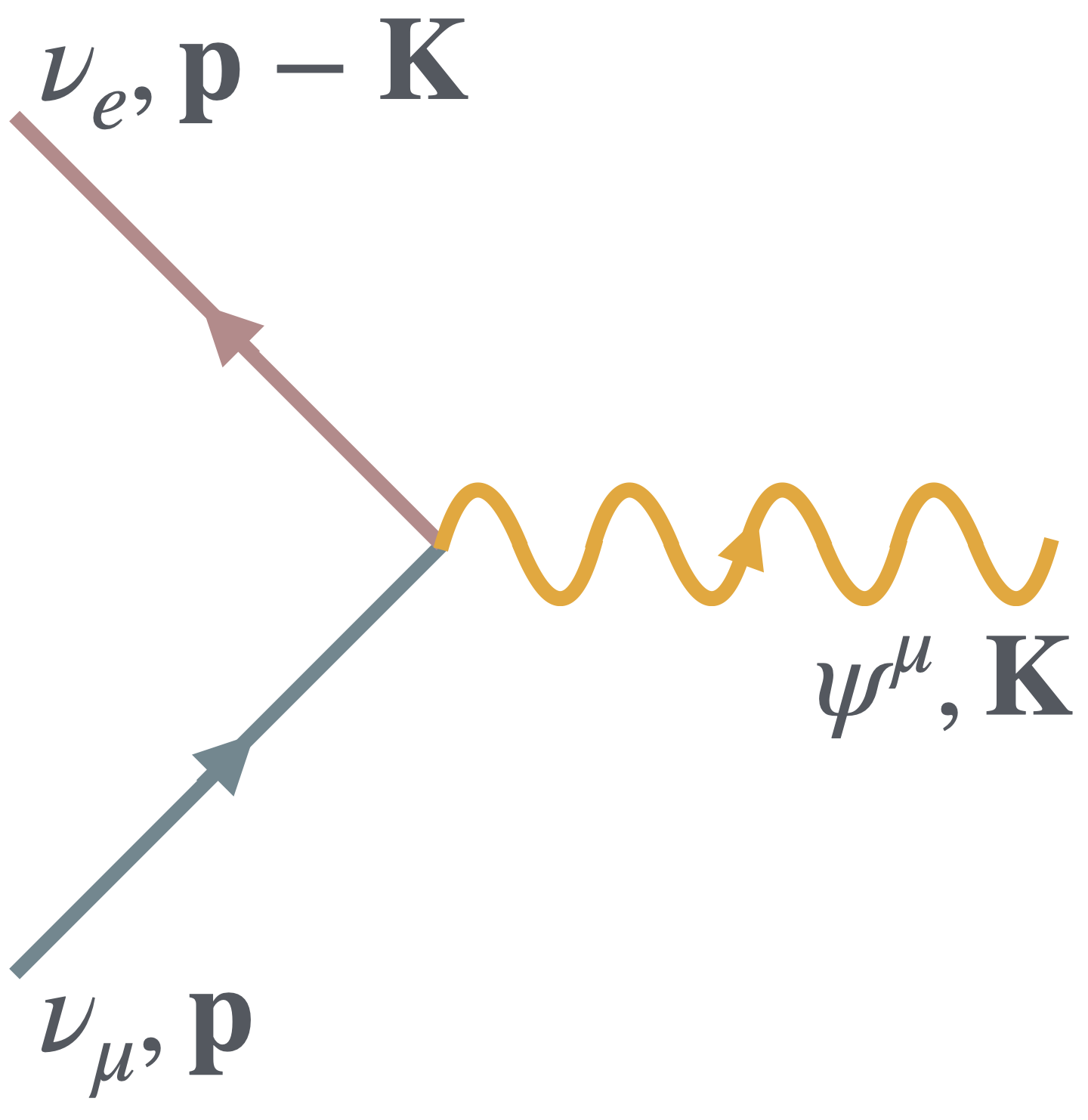}\kern3em
    \includegraphics[height=3.0cm]{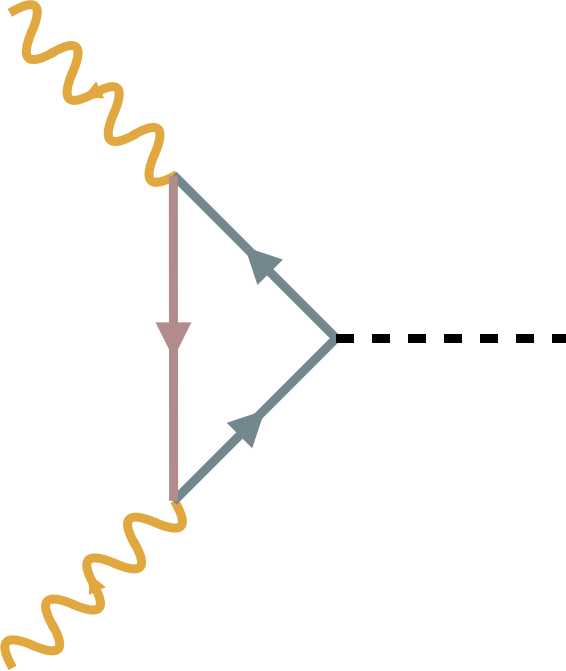}
	\caption{Quantized flavor waves (flavomons) $\Psi$ interacting with neutrinos. {\bf Left.} A $\nu_\mu$ with momentum $\bp$ decays into a $\nu_e$ with momentum $\bp-\bK$ and a flavomon of momentum $\bK$. This flavomon carries one unit of $\mu$ flavor number and minus one unit of $e$ lepton number and a very small amount of energy. {\bf Right.} A neutrino triangle diagram connects two flavomons with a neutrino plasmon. The latter is essentially a flavor-antisymmetric density wave, in which $\nu_e$ and $\nu_\mu$ densities oscillate in opposite phase so that the total density is conserved. Such processes become important beyond the linear regime of flavor evolution. (Adapted from Ref.~\cite{Fiorillo:2025npi},
    \copyright~2025 American Physical Society.)}
	\label{fig:flavomon}
\end{figure}

The theory of flavor waves as independent degrees of freedom was recently developed all the way to a kinetic equation for neutrinos together with quantized flavor waves ({\em flavomons}) \cite{Fiorillo:2025npi} that can be emitted or absorbed as sketched in Fig.~\ref{fig:flavomon}. Beyond linear theory, density waves in the neutrino flavor plasma, {\em neutrino plasmons} (dashed line in Fig.~\ref{fig:flavomon}), also appear and processes such as flavomon coalescence become possible. In principle, this approach might be able to account for flavor evolution without having to resolve the small spatial and temporal scales corresponding to flavomon wave numbers and frequencies~\cite{Fiorillo:2025npi, Johns:2025yxa}. The corresponding quasi-linear theory, in which flavomons appear as linear excitations of the plasma, and includes their non-linear feedback on the neutrino distribution, has been applied to a toy problem with discrete beams, correctly predicting the outcome of numerical simulations~\cite{Fiorillo:2024qbl}.

These ideas have not yet been developed into a practical algorithm that could be paired with a SN code. For now, it is more common to numerically solve the standard neutrino kinetic equations under various simplifying assumptions, which are difficult to validate. The most common trend is to solve them in small boxes, much smaller than the hydrodynamical resolution of SN simulations, establishing an approximate equilibrium state inferred from local conditions, which can be coupled to simulations~\cite{Bhattacharyya:2020jpj, Richers:2021nbx, Xiong:2024pue, Akaho:2025giw}. This approach was motivated by the predominance of interest in fast flavor instabilities of the last decade. These grow over short length scales of a few cm.

On the other hand, it appears now clear that slow flavor instabilities, requiring much less stringent conditions, emerge much earlier than the fast ones, already a few tens of milliseconds after core bounce \cite{Fiorillo:2025gkw}. Their slower growth casts doubts on the validity of the assumption of locality, which was already questioned in the fast flavor case in Refs.~\cite{Cornelius:2023eop, Fiorillo:2025ank}. Other approaches include artificially reducing the self-interaction strength to make the hierarchy of scales numerically manageable~\cite{Nagakura:2022kic,Nagakura:2022xwe}---but the impact of such artificial distortions cannot be self-consistently assessed---or attempting to represent the neutrino angular distribution through a few moments~\cite{Strack:2005ux, Zhang:2013lka, Grohs:2023pgq, Froustey:2023skf}. This approach does not remove the spatial hierarchy, but partially removes the computational cost; though at least for weak instabilities, it appears that a continuous representation of the angular distribution is crucial~\cite{Fiorillo:2024bzm, Fiorillo:2024uki, Fiorillo:2024pns, Fiorillo:2024dik, Fiorillo:2025ank, Fiorillo:2025zio}. Another approach is to solve the equations self-consistently while spatially under-resolving the flavor field, a method claimed to make no significant difference to the final outcome in certain numerical examples~\cite{Shalgar:2022rjj,Shalgar:2022lvv,Cornelius:2023eop,Cornelius:2024zsb}; the generality of these findings remains unclear~\cite{Nagakura:2025brr}, as small-scale flavor waves are certainly produced in fast and slow flavor instabilities~\cite{Fiorillo:2025zio}. It is difficult to do justice to the different threads of activity in a field that is so strongly in flux, with around 40 research papers appearing per year recently, many of them finding results contradicting others.

On the other hand, if the first instabilities emerge only tens of ms after collapse, largely because of the extreme $\nu_e$ dominance during the prompt $\nu_e$ phase, this characteristic feature itself will not be directly affected by such pairwise conversion effects, but of course by MSW conversion described earlier. In low-mass SNe with a particularly steep density profile, it can happen that the matter effect and neutrino-neutrino refraction together cause a spectral split of the $\nu_e$ spectrum, an ``MSW prepared spectral split'' \cite{Raffelt:2007cb, Raffelt:2007xt, Duan:2007sh, Dasgupta:2008cd}. This could be a surviving example where it is not small-scale instabilities, but slow spatial evolution along the radial direction that is responsible for collective conversion, the older version of collective flavor evolution.

If some sort of flavor equipartition between the electron and heavy-flavor neutrinos obtains during the accretion phase, the flavor-dependent features shown in Fig.~\ref{fig:NeutrinoSignal} could be largely erased, a possibility first suggested by Sawyer \cite{Sawyer:2005jk}. This equilibration would not pertain to features that depend on flavor-lepton number conservation, and so would not affect either the prompt $\nu_e$ burst or the excess $\nu_e$ flux caused by deleptonization, which however shows a strong hemispheric asymmetry due to the LESA instability (see Sec.~\ref{sec:non-spherical}). On the other hand, if one were to observe significant spectral differences, this would constrain the degree of equilibration \cite{Abbar:2024nhz, Capanema:2024hdm}. Such differences could be seen through detectors that have different sensitivities to different species and/or through Earth matter modulations.

Of course, the flavor dependence of the neutrino fluxes during the accretion phase also affects the explosion mechanism itself as it is energy deposition through beta processes in the gain region that should rejuvenate the shock wave. This question has recently been addressed in several parametric studies that represent pairwise conversion by an algorithm that attempts to mimic what might be the main features for neutrino transport, such as local flavor equilibration under the constraints of the required conservation laws \cite{Nagakura:2023mhr, Ehring:2023lcd, Ehring:2023abs, Ehring:2024mjx, Wang:2025nii, Wang:2025vbx}. The short answer is that these effects are not negligible and can modify the explosion dynamics or GW signal, and can also affect NS mergers and concomitant nucleosynthesis \cite{Wu:2017drk, Qiu:2025kgy}. In other words, quantitatively accurate SN simulations should incorporate collective flavor conversion in their neutrino-transport algorithm. 

The overall understanding of this subject remains in flux, with various conceptual and practical threads of development unfolding in parallel that will need to converge. Ideally, these strands should lead to a shared conceptual and practical framework, culminating in a numerical module that can be integrated with realistic SN simulations. However, while a complete understanding of these collective flavor effects on SN dynamics and nucleosynthesis is still lacking, it does not appear that the overarching Bethe--Wilson paradigm outlined in the previous sections will be overturned, whereas quantitative aspects of explosion dynamics, nucleosynthesis, and expected future signal characteristics remain to be settled. In particular, various signatures of collective or ordinary flavor evolution reported in older reviews~\cite{Mirizzi:2015eza}, such as spectral splits/swaps or energy modulations after signal propagation through the Earth, rely on uncertain assumptions.

\section{SN 1987A and its neutrinos}
\label{sec:SN1987A}

\subsection{The nearest supernova in centuries}

\noindent Supernova neutrinos have been observed only once, on 23~Feb\-ru\-ary 1987, when the blue supergiant Sanduleak $-69\,202$ exploded in the Large Magellanic Cloud (LMC) in the southern sky, a small satellite galaxy of our Milky Way at a distance of 50~kpc (160.000 light years), whereas, for comparison, the Galactic center is at 8~kpc. It was the closest SN in centuries. The previous historical ones of the second millennium were SN~1006 (distance 2.0~kpc), SN~1054 (2.2~kpc), that formed the Crab Nebular and Pulsar, SN~1181 (2.3~kpc), SN~1572 (Tycho's SN at 7.0~kpc), SN~1604 (Kepler's SN at 10~kpc), and Cas~A (probably Flamsteed's SN around 1680 at 3.4~kpc). Only 15--20\% of all Galactic SNe would have been visually observed, so the historical record agrees with a rate of a few per century. In addition, SN~1885A\footnote{The naming convention is that historical SNe are denoted by the prefix SN and the year. Later, when several ones were seen per year, a suffix capital A, B, \ldots, Z is added, and beyond 26 SNe per year, they are denoted by the lower-case suffix aa, ab, ac, \ldots, and after that, three lower-case letters, and so forth.} was observed in Andromeda (765~kpc), our large twin galaxy in the local group, that has produced no other SN since.

\begin{figure}[t]
\vskip4pt
	\centering
    \hbox to\columnwidth{\includegraphics[width=0.33\columnwidth]{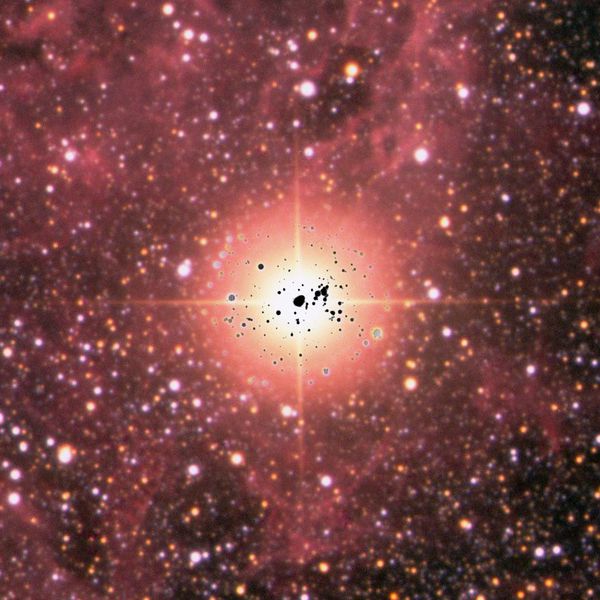}\hfill
    \includegraphics[width=0.33\columnwidth]{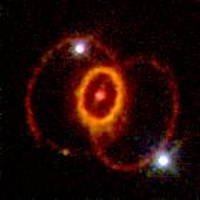}\hfill
    \includegraphics[width=0.33\columnwidth]{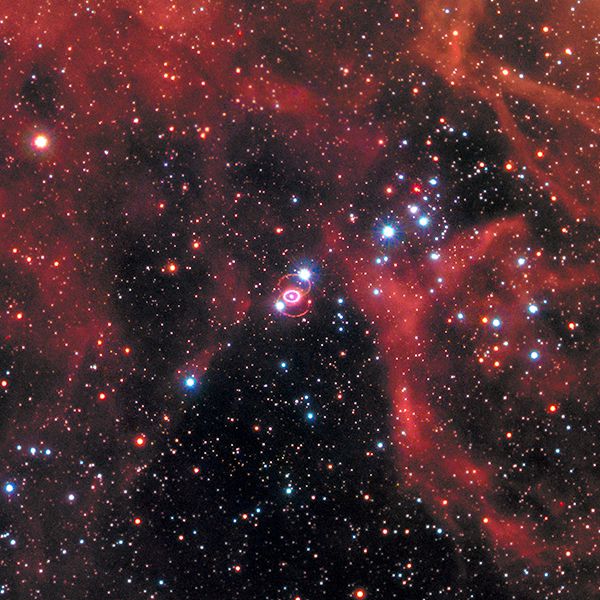}}
	\caption{SN~1987A in the Large Magellanic Cloud at a distance of 50~kpc, the explosion of the blue supergiant Sanduleak $-69\,202$ on 23~February 1987. {\bf Left.}~Supernova image on 4~March 1987, superimposed with a negative image of the pre-explosion star field, with the progenitor star at the center. (Image David Malin, \copyright\ Australian Astronomical Observatory.) {\bf Middle.} Ring system of the progenitor star, HST image from February~1994, consisting of material ejected during binary evolution of the progenitor star and illuminated by the UV flash of the explosion. SN~1987A itself is at the center of the inner ring and the two white blobs are foreground stars (\href{https://hubblesite.org/contents/media/images/1994/22/158-Image.html}{NASA/STScI}). {\bf Right.} HST image of the LMC, with SN~1987A at the center, taken in January~2017, 30~years after the explosion (\href{https://hubblesite.org/contents/media/images/2017/08/3987-Image.html?page=1&keyword=1987A}{NASA/STScI}).}
	\label{fig:SN1987A}
\end{figure}

After the first visual discovery on 24 February 1987, signs of optical brightening were found on plates taken by McNaught at 10:39 UT (universal time) on 23 February 1987, and based on the first neutrino in the IMB detector (see below), the explosion occurred around 3 hours earlier at UT 7:35:41.374 $\pm$ 50~ms. To cite a paper written on occasion of the 20th anniversary \cite{Fransson:2007}: ``The unique supernova SN 1987A has been a bonanza for astrophysicists. It provided several observational ‘firsts’, like the detection of neutrinos from the core collapse, the observation of the progenitor star on archival photographic plates, the signatures of a nonspherical explosion and mixing in the ejecta, the direct observation of supernova nucleosynthesis, including accurate masses of $^{56}$Ni, $^{57}$Ni and $^{44}$Ti, observation of the formation of dust in the supernova, as well as the detection of circumstellar and interstellar material.'' 

The left panel of Fig.~\ref{fig:SN1987A} shows a negative-image superposition of the star field around the SN, revealing at the center the previously known blue supergiant Sanduleak $-69\,202$, the first time a SN explosion could be correlated with a specific progenitor star. It was not the usual red supergiant progenitor---in Sanduleak's case, it must have undergone a binary evolution that produced the three-ring structure of interstellar material visible in the middle panel. This was one of the first images taken by the repaired Hubble Space Telescope (HST) in February 1994, and today is an emblematic image of SN~1987A. The SN itself is the bright spot at the center, here seven years after the explosion. It expands and began interacting with the inner ring (radius 245 light-days) some 20 years after the explosion.

However, until today, nearly 40~years after the explosion, no NS was found in the remnant of SN~1987A, sometimes stoking speculations that a BH might have formed after all \cite{Blum:2016afe, Bar:2019ifz}. In particular, there is no evidence for a pulsating signal, but of course, not all NSs become pulsars. However, in 2019 the Atacama Large Millimeter/submillimeter Array (ALMA) presented high angular resolution images of dust and molecules in SN 1987A ejecta that reveal a localized blob of warm dust, possibly corresponding to a compact source with a luminosity of 40--$90\,L_\odot$ as explained in Ref.~\cite{Page:2020gsx}. Likewise, recent near- and mid-infrared integral field spectroscopy with the James Webb Space Telescope (JWST) can be explained by ionizing radiation from a NS illuminating gas from the inner parts of the exploded star~\cite{Fransson:2024csf}. The displacement of the hot blob from the SN center is consistent with a transverse NS kick velocity of 300--$500~{\rm km}~{\rm s}^{-1}$ at~birth~\cite{Page:2020gsx}, similar to an estimate based on the JWST observations~\cite{Larsson+2025}.

\subsection{Neutrino detection}

\label{sec:SN1987Adetection}

\noindent However, the neutrino observations were the main sensation relevant for this review. At the time of SN~1987A, four detectors were sensitive to the neutrino burst and registered a total of two dozen events. By far the dominant detection process was IBD ($\overline\nu_ep\to ne^+$), although one cannot exclude that elastic electron scattering $\nu e^-\to e^- \nu$ could have accounted for perhaps one of the events. Either way, the final-state $e^\pm$ with energy of a few 10~MeV can be seen by the Cherenkov or scintillation light produced within the detector volume. The thriller of identifying the signal in the different detectors is recollected, for example, by Masatoshi Koshiba \cite{Koshiba:1992yb}, who received the 2002 Nobel Prize in Physics in part for this discovery. The data and detectors were recently reviewed from a contemporary perspective \cite{Fiorillo:2023frv} and in the following we mainly draw from this exposition.

\begin{figure}[h]
	\centering
    \includegraphics[width=\columnwidth]{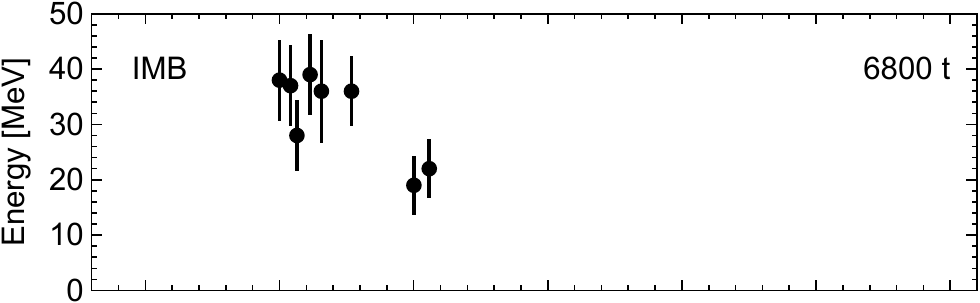}
    \includegraphics[width=\columnwidth]{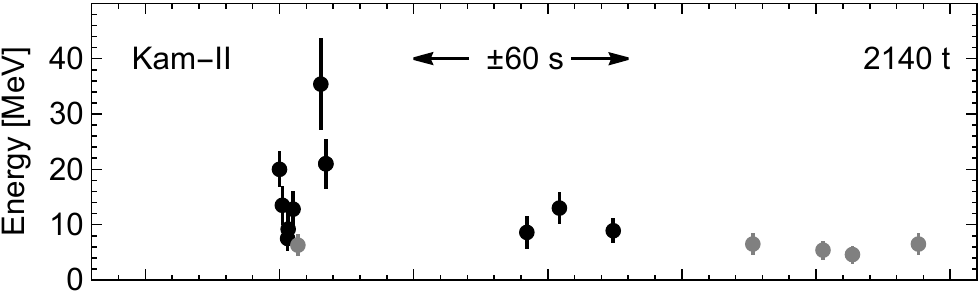}
    \includegraphics[width=\columnwidth]{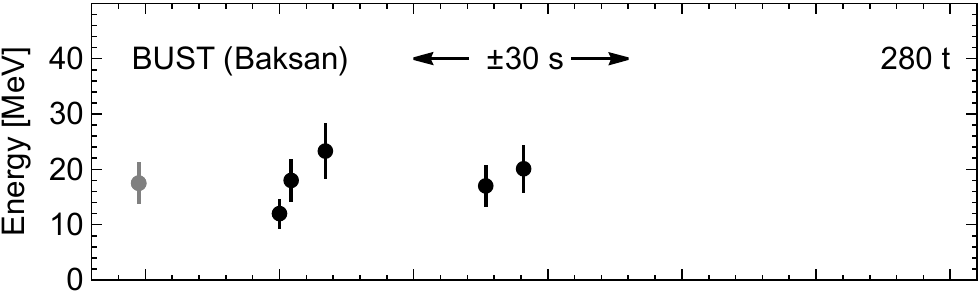}
    \includegraphics[width=\columnwidth]{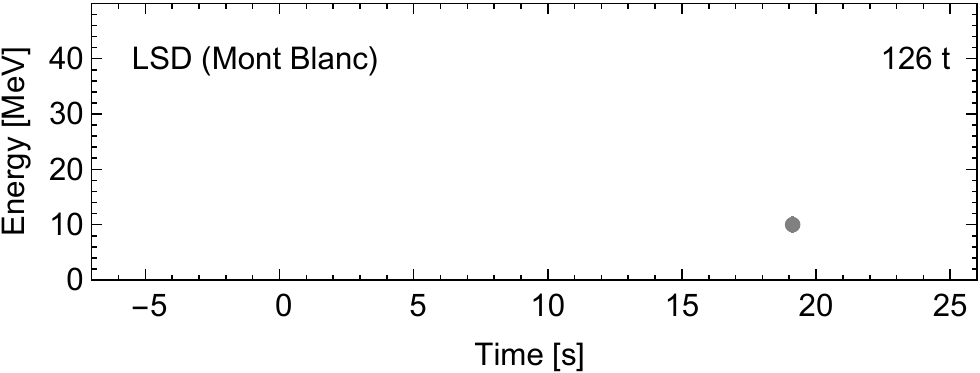}
	\caption{Neutrino signal of SN 1987A at IMB, Kam-II, BUST and LSD. We show the detected positron energy as a function of time relative to the first IMB event  (7:35:41.374 UT on 23 February 1987). The first events of Kam-II and BUST were aligned with that of IMB because both detectors had a significant clock uncertainty. For BUST (Baksan), we also show an earlier event that was attributed to background. LSD measured an additional burst nearly five hours earlier not shown here and no corresponding events in the other detectors. In Kam-II, event No.~6 (with lowest energy in the main burst) and those after 17~s as well as the late LSD events (another one at 37.5 s) are usually attributed to background, here shown in gray. The fiducial masses of the scintillator detectors BUST and LSD are water equivalents, i.e., normalized to the number of protons. (Adapted from Ref.~\cite{Fiorillo:2023frv}, \copyright\ 2023 American Physical Society.)}
	\label{fig:SN1987A-Neutrinos}
	\vskip-4pt
\end{figure}

The largest detector was the Irvine-Michigan-Brookhaven (IMB) water Cherenkov detector \cite{Bionta:1987qt, IMB:1988suc}, located in the Morton-Thiokol salt mine (Fairport, Ohio, USA), with a sensitive mass of 6800~t, but relatively sparse photo sensor coverage and corresponding high energy threshold. The energies and time sequence of registered events is shown in the top panel of Fig.~\ref{fig:SN1987A-Neutrinos}. IMB had no low-energy background for the short SN signal duration. This detector had been built to search for proton decay, a phenomenon not yet found today even with much larger detectors, and IMB itself was decommissioned around 1991.

The second largest experiment was the Kamiokande-II (Kam-II) water Cherenkov detector with 2140~t sensitive mass, but much lower threshold, making it competitive for SN neutrino detection \cite{Kamiokande-II:1987idp, Hirata:1988ad}. It had also been built to search for proton decay (Kamiokande = Kamioka Nucleon Decay Experiment) and is located in the Mozumi Mine, Kamioka Section of Hida, Gifu Prefecture, Japan. After not finding proton decay, the experiment was refurbished to reduce low-energy backgrounds, and moved on to a long episode (until 1995) of solar neutrino measurements. (In 1996, the much larger Super-Kamiokande water Cherenkov detector took up operation and the original site is now occupied by the KamLAND experiment, a 1~kt scintillator detector that measured reactor and geophysical neutrinos and today searches for neutrinoless double beta decay.) SN~1987A occurred within months after implementing the original improvements. The time sequence of SN~1987A events is shown in the second panel of Fig.~\ref{fig:SN1987A-Neutrinos} such that the first events in IMB and Kam-II coincide. IMB had good absolute clock time, whereas that of Kam-II was uncertain to $\pm1$~min so that the relative timing is totally uncertain within the burst duration of a few seconds. Besides atmospheric muons at large energies, Kam-II had a low-energy background mostly from the beta decay of $^{214}$Bi with a rate of 0.187~Hz. Based on an energy cut, event No.~6 is usually attributed to background, just like the very late events after 17~s. According to these assumptions, there are 11 SN-related events with an average positron energy of 15.4~MeV, but a conspicuous and often-discussed time gap between a first bunch of events until 2~s, and three late~events~after~9~s.\looseness=1

The third experiment to observe SN~1987A neutrinos was the Baksan Scintillator Underground Telescope (BUST), operated by the Institute of Nuclear Research in Moscow \cite{Alekseev:1987ej, Alekseev:1988gp}. This detector, with a water-equivalent mass of 280~t, is located in the Baksan underground laboratory under Mount Andyrchi in the North Caucasus. The tunnel entrance is near the village ``Neutrino'' in the Baksan valley. BUST consists of 3150 separate elements of dimension 70${\times}$70${\times}$30~cm, filled with scintillator and viewed by one photomultiplier. BUST has surveyed the Galaxy for neutrino bursts since 30~June 1980 and is still operating today, but no signal was observed other than SN~1987A \cite{Alekseev:1993dy, Novoseltsev:2022lmd}. Around the time of SN~1987A, a bunch of six events was found, shown in the third panel of Fig.~\ref{fig:SN1987A-Neutrinos}, also with significant clock uncertainty. The second event is lined up with the first event in IMB, whereas the first BUST event was attributed to background. The overall large background rate implies that such a 5-event burst could randomly occur perhaps once per day, but the temporal correlation with IMB and Kam-II suggests a SN~1987A origin, although one would have expected only around 1 SN-related event, given the small size of the detector.

The fourth and smallest detector was the Liquid Scintillator Detector (LSD) in the Mont Blanc tunnel (between France and Italy) with a water-equivalent mass of 126~t, but better trigger efficiency than BUST, so the expected SN signal is about 2/3 that of BUST, i.e., one might have expected perhaps 1~event. This detector observed a burst almost five hours earlier than the other detectors, consisting of 5~events within 7~s, which could not be related to SN~1987A in any plausible way, even though LSD was first to announce a SN~1987A related neutrino signal \cite{Dadykin:1987ek, Aglietta:1987it}. On the other hand, LSD was of course also sensitive at the burst time of the other detectors and it had good absolute clock time. As shown in the bottom panel of Fig.~\ref{fig:SN1987A-Neutrinos}, it observed an event at around 19~s (and another one at 37.5~s) that should be attributed to background. LSD had a low-energy background rate of 0.012~Hz, so over 13~s one expects 0.16~events. Its nonobservation during the signal at the other detectors provides a constraint on the overall neutrino flux, which however is not very restrictive. LSD was built to search for a Galactic SN burst, but  too small for a SN in the LMC. The early burst, with no signal in any of the other detectors and no straightforward SN connection, remains forever puzzling. No similar high-multiplicity event was found during the entire LSD operation that began around 1984 and ended with the devastating fire in the Mont Blanc Tunnel~on~24~March~1999.\looseness=1

\subsection{Interpretation}

\begin{figure}
    \vskip4pt
	\centering
    \hbox to\columnwidth{\includegraphics[height=0.48\columnwidth]{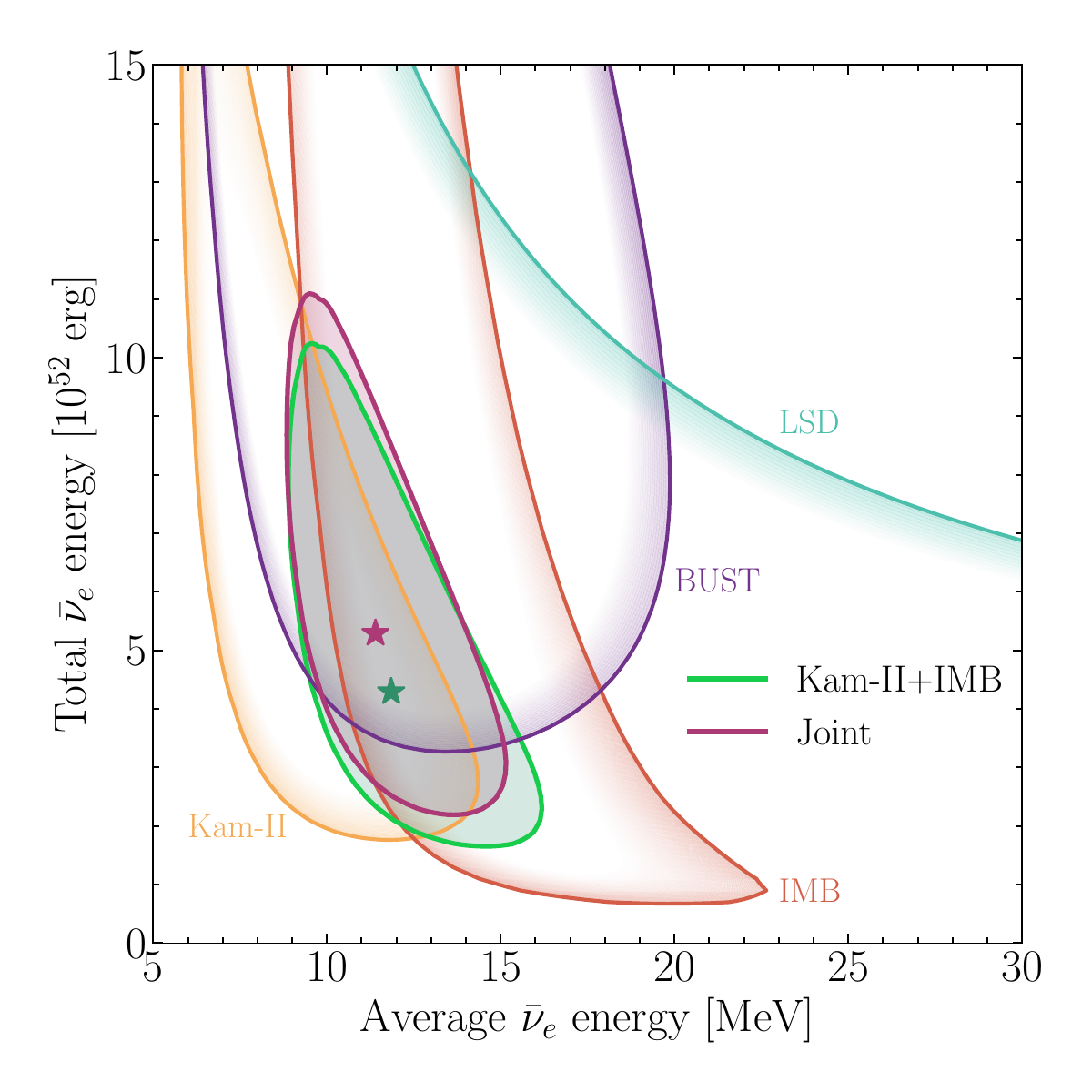}\hfill
    \includegraphics[height=0.48\columnwidth]{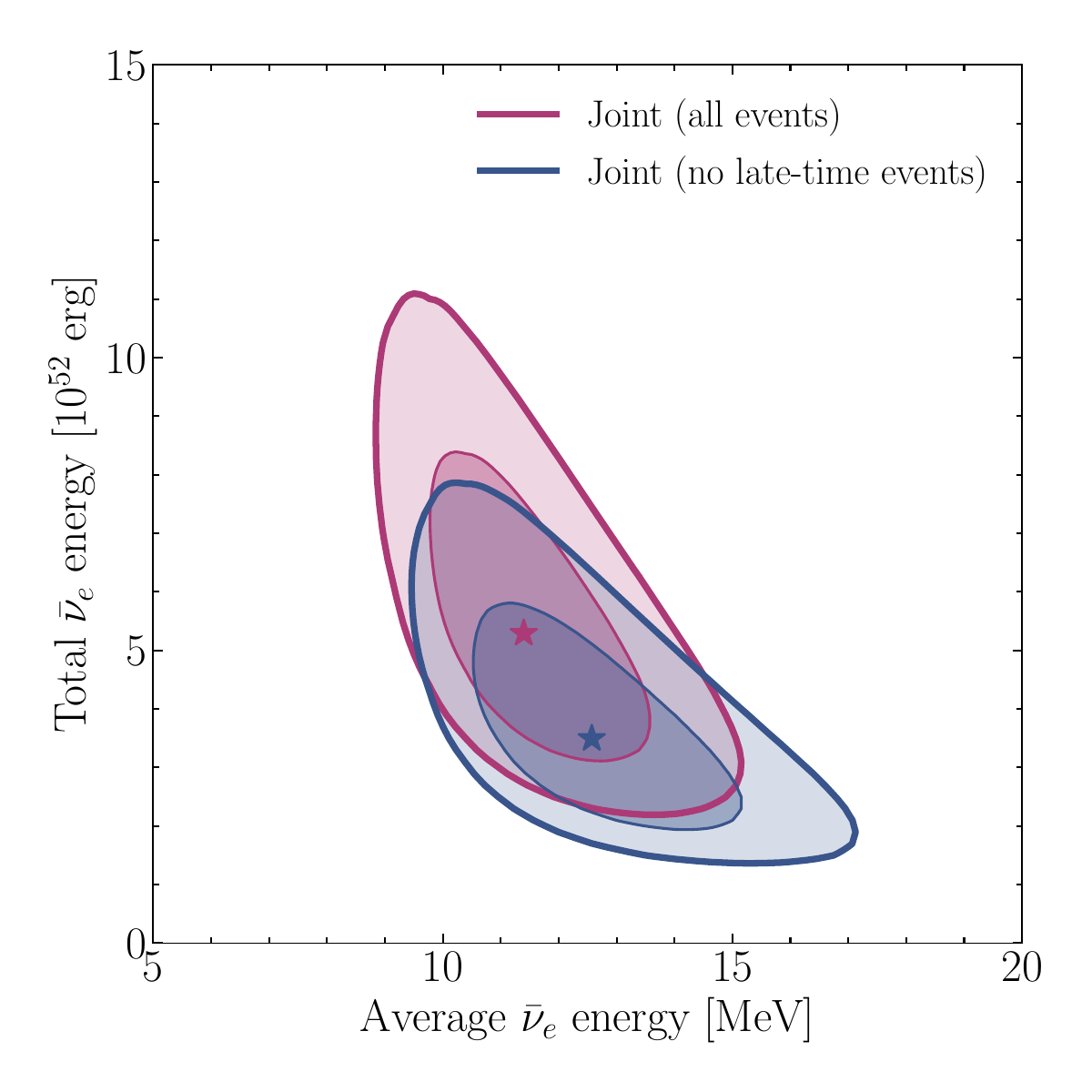}}
    \vskip4pt
	\caption{Allowed regions (95\% confidence level) for the total energy emitted by SN~1987A in the form of $\overline\nu_e$ and $\langle \epsilon_{\overline\nu_e}\rangle$, the average $\overline\nu_e$ event energy \cite{Fiorillo:2023frv}. It is assumed that the $\overline\nu_e$ flux has a quasi-thermal spectrum with a Maxwell-Boltzmann distribution with $T=\langle \epsilon_{\overline\nu_e}\rangle/3$. {\bf Left}. All experiments separately and joint analysis including all of them (magenta) or only IMB and Kam-II (green). {\bf Right}. Blow-up of the best-fit region from all experiments and all events (magenta), or leaving out the late-time events in Kam-II and BUST (blue) that might have another origin than proto-neutron star cooling as discussed in the text. (Figures from Ref.~\cite{Fiorillo:2023frv}, \copyright\ 2023 American Physical~Society.)
    }
	\label{fig:SN1987A-Parameters}
	\vskip4pt
\end{figure}

\noindent The statistics of these observations is quite limited; a very detailed interpretation is not supported by such sparse data. One common approach is to model the time-integrated signal as a Maxwell-Boltzmann distribution for the $\overline\nu_e$ that were presumably detected through IBD. In this case, the only open parameters are the average $\overline\nu_e$ energy and the total energy emitted in the form of $\overline\nu_e$. The parameter regions allowed at 95\% confidence level by a maximum likelihood analysis are shown in Fig.~\ref{fig:SN1987A-Parameters} for all experiments separately, and a joint analysis (magenta). A joint analysis using only the IMB and Kam-II data is shown in green, the results being quite similar, so including or excluding the small detectors makes no substantial difference, but of course, there is no objective reason to leave them out. Within the large uncertainties caused by the scarcity of data one finds $\langle E_{\overline\nu_e}\rangle\simeq11$--$12~{\rm MeV}$ and a total emitted energy of around $3\times10^{53}~{\rm erg}$ (six times the total $\overline\nu_e$ energy), so the broad-brush picture agrees perfectly with~expectations.\looseness=1

\begin{figure*}
\centering
\includegraphics[width=0.89\textwidth]{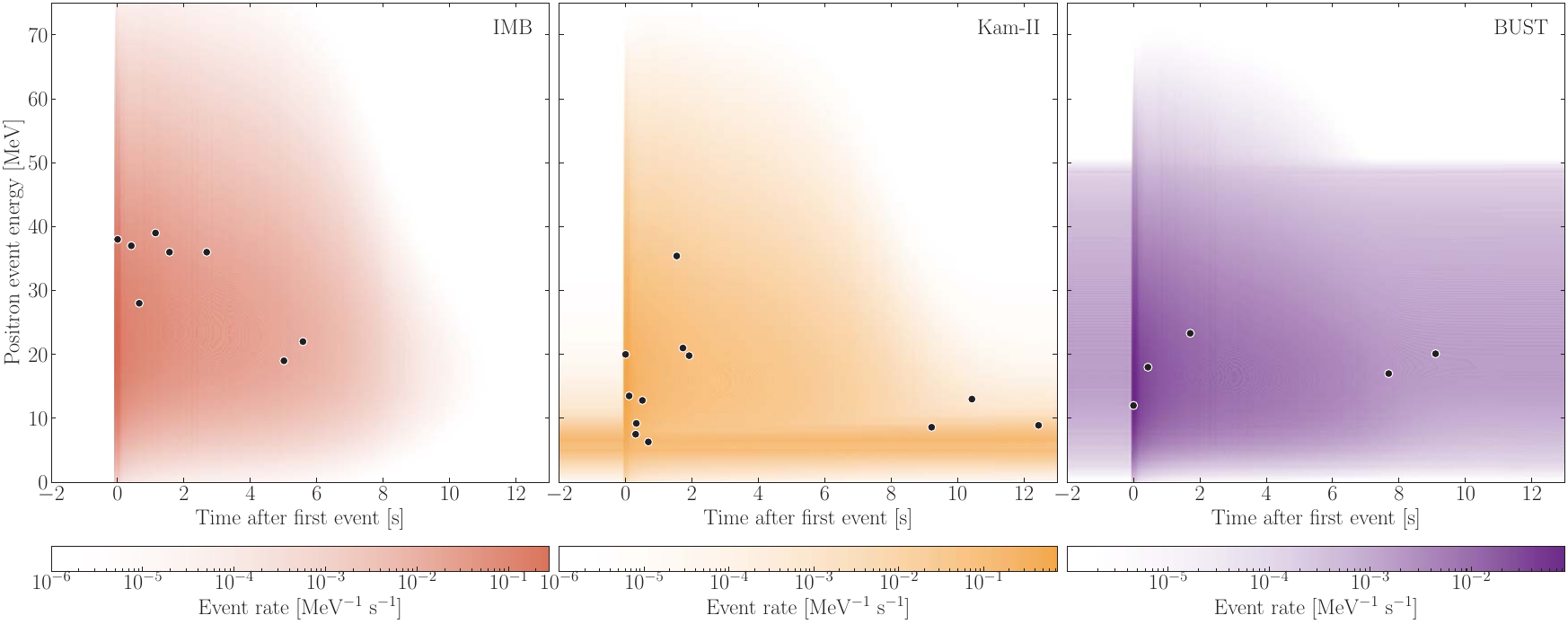}
\caption{Differential event distribution (signal and background) at each experiment, compared with the observations. Results are shown for model 1.44-SFHo, the best fit case of Ref.~\cite{Fiorillo:2023frv}, without neutrino flavor swap; each offset time is chosen as the best-fit value. The main conclusion is that the late events in Kam-II and BUST are not really compatible with the much shorter cooling time of the numerical models which is caused by convection in the proto-neutron star that accelerates energy and lepton-number loss compared with diffusive transport alone. (Figure from Ref.~\cite{Fiorillo:2023frv}, \copyright\ 2023 American Physical~Society.)
}\label{fig:2DDistributions}
\vskip-2pt
\end{figure*}

More detailed interpretations have gone in two different directions. One is to fit the parameters of certain assumed analytical emission models \cite{Loredo:2001rx, Bozza:2025wqo}, that are motivated by the standard SN paradigm, but not calibrated against numerical simulations. On the other hand, one should perform a direct comparison with modern numerical simulations, depending on the assumed progenitor model, nuclear equation of state, and assumptions about neutrino flavor evolution. Such an exercise was performed recently \cite{Fiorillo:2023frv} and once more it emerges that data and models broadly agree, but a detailed discrimination is difficult, in part because of the intrinsic tension between the higher-energy IMB and lower-energy Kam-II data. Indeed, the average energy of the Kam-II events in the first second is particularly low, leading some authors to conclude that this effect, relative to a large suite of numerical models from the literature, was a significant problem for contemporary SN theory \cite{Li:2023ulf}. However, the significance of this tension, when accounting for the ``look elsewhere effect'', does not even reach the $1\sigma$ level \cite{Fiorillo:2023frv}, so the scarce statistics prevents a conclusive~interpretation.\looseness=1

Another and probably more significant issue is a tension of the overall signal duration between models and data. All simulations produce relatively short signals when the generic effect of PNS convection is included that must operate during the few-second cooling period after core bounce. These short predicted signal durations are seen in Fig.~\ref{fig:2DDistributions}, comparing the data with the time-energy distribution predicted by the best-fit model of Ref.~\cite{Fiorillo:2023frv}. It looks as if the late-time events in Kam-II (the 3-event cluster after 9~s) and the two late events in BUST might have a different cause. They are difficult to account for by background, but SN-related effects are also not straightforward. These could include the late accretion of material returned by spherical fallback~\cite{Akaho:2023alv}, or more likely by highly asymmetric, large-scale convective downflows~\cite{Janka:2025tvf}. Another possibility is a sudden NS contraction and reheating by a QCD phase transition \cite{Takahara+1988, Gentile:1993ma, Sagert:2008ka, Fischer+2018, Kuroda+2022, Zha:2021fbi, Jakobus:2022ucs, Huang:2024xff, Largani:2023oyk}. If such effects could really produce the late signal warrants further investigation. Certainly, these questions highlight the under-appreciated importance of the very late neutrino signal of the next Galactic SN \cite{Dasgupta:2009yj, Li:2020ujl}.

\section{New particles}
\label{sec:New-Particles}

\noindent The idea that new feebly interacting particles---or neutrinos with nonstandard properties---could affect SN physics dates back to a paper by Falk and Schramm (1978) \cite{Falk:1978kf}, who argued that neutrino radiative decays $\nu\to\nu'+\gamma$ should not deposit excessive energy within the progenitor star, so as to avoid overly energetic SN explosions. Later, the SN~1987A neutrino observations motivated numerous constraints on new particles---most notably axions \cite{Raffelt:1987yt, Turner:1987by, Ellis:1987pk, Mayle:1987as, Ericson:1988wr, Mayle:1989yx, Brinkmann:1988vi, Burrows:1988ah, Burrows:1990pk, Turner:1991ax, Raffelt:1993ix, Janka:1995ir, Keil:1996ju, Hanhart:2000ae, Fischer:2016cyd, Chang:2018rso, Carenza:2019pxu, Carenza:2020cis, Fischer:2021jfm, Choi:2021ign, Betranhandy:2022bvr, Lucente:2022vuo, Vonk:2022tho, Ho:2022oaw, Li:2023thv, Lella:2023bfb, Springmann:2024ret, Mori:2025cqf, Fiorillo:2025gnd}---that could carry away too much energy directly from the SN core. Such new energy-loss channels would shorten the neutrino burst, whose duration reflects the SN cooling speed. From the observed SN~1987A neutrino burst duration of a few~seconds, a schematic bound is~\cite{Raffelt:1990yz, Raffelt:2006cw},
\begin{equation}
    \epsilon_x<10^{19}~{\rm erg}~{\rm g}^{-1}~{\rm s}^{-1},
\end{equation}
where the new energy-loss rate $\epsilon_x$ is to be evaluated at typical SN core conditions of $\rho=3\times10^{14}~{\rm g}~{\rm cm}^{-3}$, $T=30$~MeV, and proton abundance of $Y_p=0.15$, representative of conditions at around 1~s pb. Of course, this criterion is only indicative for a quick estimate, not a precision bound, notably in view of the scarce SN~1987A data. 

\begin{figure}[t]
\centering
\includegraphics[width=1.0\columnwidth]{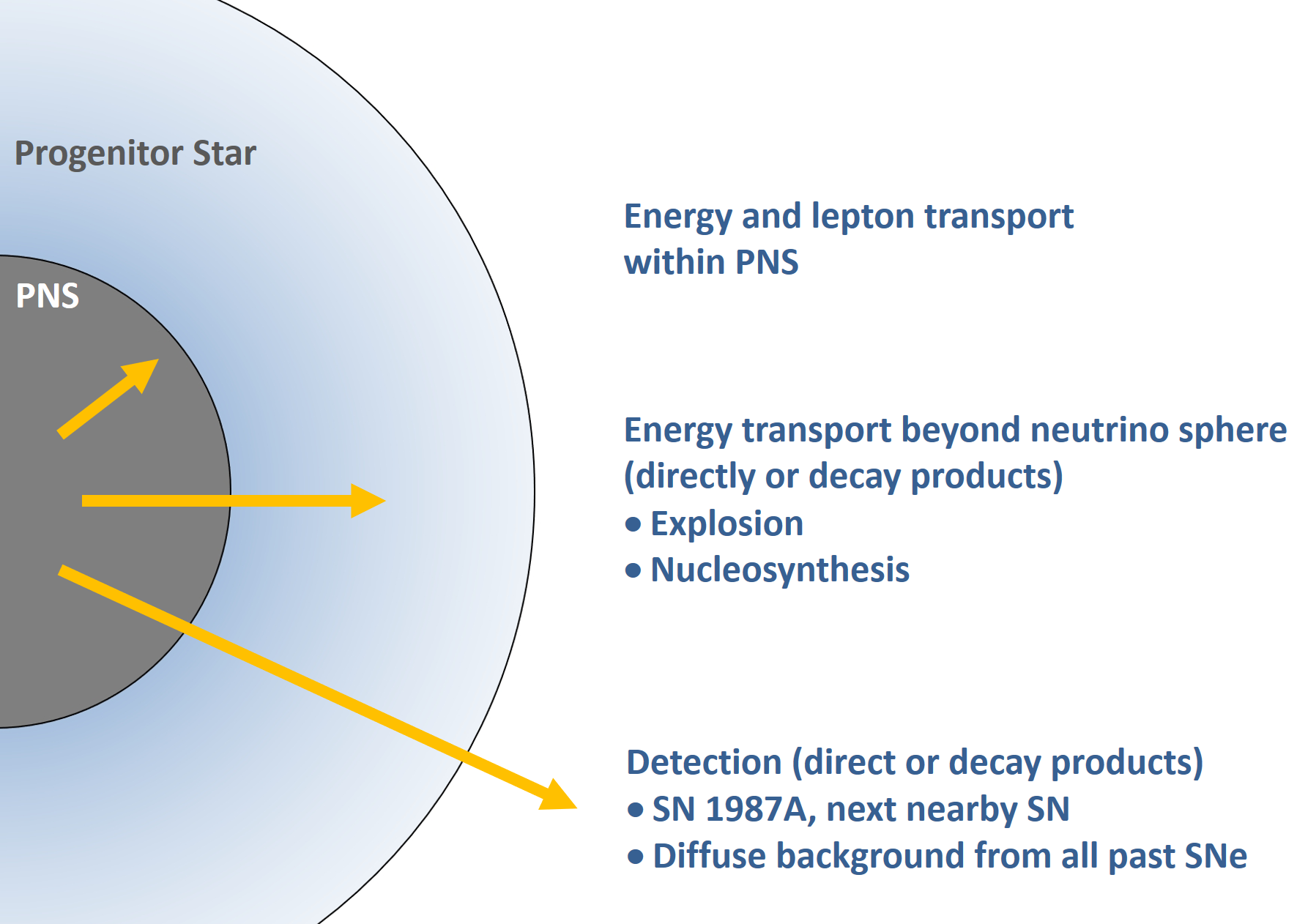}
\caption{New particles other than standard neutrinos can affect SN physics in various ways as discussed in the text. They can transport energy within the SN or simply remove it and thereby modify SN physics or they can become visible by decays or conversion to visible particles, either from the historical SN~1987A, the next Galactic SN, or from all past SNe in the universe.}\label{fig:NewParticles}
\end{figure}

New feebly interacting particles, depending on their exact properties, can do everything in a SN that neutrinos can do: carry away or transport energy and deposit energy within the progenitor star, thus enhancing the explosion energy, or becoming visible, either directly in detectors, or through decays, or by conversion such as magnetically induced axion-to-photon oscillations. Such arguments pertain to the historical SN~1987A, the next nearby SN, or all past SNe in the universe in analogy to the DSNB---see Fig.~\ref{fig:NewParticles} for a schematic representation of the different arguments. Analogous arguments can also pertain to the NS remnants that could cool too fast \cite{Leinson:2021ety, Beznogov:2018fda, Buschmann:2021juv, Gomez-Banon:2024oux,Fiorillo:2025zzx}, could show excess radiation from axion-photon conversion \cite{Morris:1984iz, Raffelt:1987im, Fortin:2018ehg, Dessert:2019dos, Buschmann:2019pfp}, or suffer from a modified equation of state by axion-field backreaction \cite{Hook:2017psm, Balkin:2023xtr, Kumamoto:2024wjd}.

Early reviews of astrophysical particle constraints can be found in Refs.~\cite{Raffelt:1990yz, Raffelt:1996wa, Raffelt:2006cw}, while more recent updates focusing specifically on axions are \cite{Caputo:2024oqc, Carenza:2024ehj}. Over the past few years, there has been a resurgence of interest in low-mass particles and the associated astrophysical bounds, leading to an overwhelming number of new papers. Still, aside from the cited works on axions, there is currently no comprehensive and up-to-date review. If any of these particles---such as axions, sterile neutrinos, or scenarios involving nonstandard neutrino properties---were to be experimentally discovered, they would almost certainly impact compact transient astrophysical phenomena, including SNe and NS mergers. At present, however, SN physics is primarily used to constrain the parameter space of such models. For illustration, we now highlight a few representative cases.

{\bf Neutrino masses.}---Concerning neutrinos themselves, relative to the early days of SN physics, flavor conversion through mass--mixing is now firmly established, but its role in SN physics remains a subject of active research (Sec.~\ref{sec:FlavorEvolution}). The neutrino masses themselves, on the other hand, have no tangible impact. If they are of Dirac type, sterile partners of the active species exist and in any elastic collision, they would emerge with a relative probability of about $(m_\nu/\epsilon_\nu)^2\sim10^{-16} m_{\rm eV}^2$, assuming $\epsilon_\nu=100$~MeV. As estimated earlier, a neutrino with this energy scatters around $10^9$ times during PNS cooling, so at least one sterile partner is produced for $m_\nu\gtrsim 3$~keV. Such questions were of particular interest in the early 1990s, when experimental signatures suggested a 17~keV neutrino and many authors derived SN~1987A cooling bounds of $m_{\rm Dirac}\lesssim 30$~keV \cite{Raffelt:1987yt, Gaemers:1988fp, Grifols:1990jn, Burrows:1992ec, Raffelt:1996wa}, which today are mostly of historical interest. The latest experimental limit on the overall neutrino mass scale from the KATRIN experiment is $m_\nu<0.45$~eV (90\% CL) \cite{KATRIN:2024cdt}, whereas cosmological limits with different assumptions are in the range $\sum m_\nu<0.1$--0.5~eV \cite{ParticleDataGroup:2024cfk}. Either way, the SN production of sterile Dirac partners is strongly suppressed.

{\bf Lepton-number violation.}---If neutrino masses were of Majorana type, the helicity-flipping reactions would produce (active) antineutrinos instead of sterile neutrino partners, but again, the scattering rate is suppressed by the same chirality factor of around $(m_\nu/\epsilon_\nu)^2$. Therefore, in a SN core, lepton number is conserved  with excellent precision, even if neutrinos are of Majorana type. In principle, lepton number could be violated by other beyond-standard-model (BSM) particle physics effects and could deleptonize internally, for instance by neutrino-majoron interactions, quickly running down the $\mu_e$ and $\mu_{\nu_e}$ chemical potentials. If this effect were rapid enough to happen already during the initial collapse, the SN core would heat up early and might produce a thermal bounce at subnuclear density \cite{Kolb:1981mc, Dicus:1982dk, Fuller:1988ega, Rampp:2002kn, Suliga:2024nng}. While such scenarios have not been systematically studied, it is intriguing that they do not necessarily prevent SN~explosions. 

{\bf Time-of-flight effects.}---If the SN explodes at a distance $D$ from Earth, neutrino masses cause the time of flight to be extended by the amount
\begin{equation}
    \Delta t=\frac{1}{2}\left(\frac{m_\nu}{\epsilon_\nu}\right)^2 D =5.14~{\rm ms}\,\left(\frac{m_\nu}{1~{\rm eV}}\right)^2 \left(\frac{10~{\rm MeV}}{\epsilon_\nu}\right)^2
    \left(\frac{D}{10~{\rm kpc}}\right).
\end{equation}
For SN~1987A at a distance of 50~kpc and a burst duration of 10~s, a rough mass bound is 20~eV, the exact value being hotly debated at the time \cite{Loredo:1988mk}. For the next nearby SN, assuming a distance of 10~kpc and if one were to observe the SASI modulations, the time-of-flight sensitivity could be $m_\nu\lesssim 0.14$~eV \cite{Ellis:2012ji}. If one were to observe the sudden signal shutdown by BH formation, the sensitivity would be a few~eV~\cite{Beacom:2000qy}. Other studies using fast signal features include Refs.~\cite{Totani:1998nf, Nardi:2004zg, Pompa:2022cxc, Denton:2024mlb}. Conversely, the cosmological bounds imply that such fast features would not be smeared out by time-of-flight delays: neutrino masses are probably not a concern for measuring fast signal variations.

A hypothetical small electric neutrino charge would deflect neutrinos in the galactic $B$ field, once more causing an energy-dependent delay of arrival times, providing a SN~1987A limit of $|e_\nu/e|\lesssim 3\times10^{-17}$ \cite{Barbiellini:1987zz}. Another hypothetical form of energy-dependent velocity variation in the generic form $v=1\pm (E/M_{\rm LV})^n$ could arise from Lorentz violation and has been constrained by SN~1987A \cite{Fujiwara:1989xf, Ellis:2008fc} or could be constrained from fast signal variations in the next SN \cite{Ellis:2011uk, Moura:2023xba}.

One may also compare the arrival times of photons with neutrinos a few hours earlier. For SN~1987A, the Shapiro time delay in the Galactic gravitational potential was estimated to be a few months, proving that neutrinos indeed gravitate because their Shapiro time delay was the same as for photons to better than at least $4\times10^{-3}$ \cite{Longo:1987gc, Krauss:1987me}. A similar argument was recently used for the arrival times of GWs and photons from the binary merger GW170817 and concomitant gamma-ray burst GRB170817A \cite{LIGOScientific:2017ync} within 1.7~s, proving that GWs had the same delay as photons to within $10^{-7}$  \cite{Shoemaker:2017nqv}, proving that GWs themselves indeed gravitate. A three-way comparison might become possible when GWs, neutrinos, and photons are observed from the next Galactic SN, as well as a hypothetical delay between $\nu$ and~$\overline\nu$.

For a brief period in 2011, it was reported that neutrinos in the CERN to Grans Sasso beam arrived faster than the speed of light, a result soon recognized as a hardware problem \cite{OPERA:2011ijq}. However, it led to a huge debate, including SN~1987A neutrinos as proof of their correct propagation speed. In the citation record of SN~1987A neutrinos, this episode shows up as a distinct spike, see \href{https://inspirehep.net/literature/244946}{https://inspirehep.net/literature/244946} for Kamiokande-II and \href{https://inspirehep.net/literature/245093}{https://inspirehep.net/literature/245093} for~IMB.

{\bf Neutrino dipole moments.}---Neutrino masses inevitably imply magnetic and/or electric dipole and transition moments \cite{Giunti:2014ixa, Giunti:2024gec}. For a given mass eigenstate $m_\nu$, the dipole moment is $\mu_{\nu}=3e\GF m_{\nu}/8\sqrt2 \pi^2=3\times10^{-19}\mu_{\rm B}m_{\rm eV}$ in units of the Bohr magneton $\mu_{\rm B}=e/2m_e$, whereas transition moments---connecting different mass eigenstates---are orders of magnitude smaller. The most restrictive limit of \hbox{$|\mu_{ij}|<1.5\times10^{-12}\mu_B$} comes from plasmon decay in the brightest red giants in globular clusters \cite{Capozzi:2020cbu}. Transition moments allow for radiative decay \hbox{$\nu\to\nu'\gamma$} or conversion between different mass eigenstates in strong electric or magnetic fields, including spin-flavor transitions. In the Majorana case, this includes effectively lepton-number violating $\nu\to\overline\nu$ transitions. 

However, the mass-induced dipole moments are far too small to engender any important effects in SNe, just like the masses themselves are too small as discussed earlier. For much larger BSM values, the picture changes. However, now the parameter space of mass-mixings, dipole moments, and $B$-field distributions is so vast, not to mention collective flavor evolution or helicity evolution in the intervening galactic $B$--field, that no generic results can be reported here. In the Dirac case, spin-flip scattering leads to the production of the sterile partners that carry away energy, leading to a SN~1987A cooling bound of a few $10^{-12}\,\mu_{\rm B}$ \cite{Barbieri:1988nh, Notzold:1988kz, Ayala:1998qz, Kuznetsov:2009zm}, comparable to the globular cluster bound.

{\bf Nonstandard interactions.}---Another possible phenomenological modification consists of neutrino nonstandard interactions (NSI) or nonstandard neutrino self-interactions (NSSI), meaning that at low energies, there are possible small modifications of neutral or charged-current interactions with other fermions or among neutrinos \cite{Proceedings:2019qno}. In the SN context, they would impact the matter effect for flavor conversion \cite{Esteban-Pretel:2007zkv, Fogli:2002xj, Stapleford:2016jgz, Jana:2024lfm, Das:2025zts, Dutta:2025rxh} or modify collective flavor evolution \cite{Blennow:2008er, Das:2017iuj, Dighe:2017sur, Yang:2018yvk, Abbar:2022jdm}. In view of the general state of flux of our understanding of collective flavor evolution in SNe, the true role of NSI and NSSI is probably hard to judge at present. Laboratory bounds are surprisingly weak on some NSI parameters, and neutrino-neutrino interactions cannot be probed experimentally at all.

{\bf Secret neutrino interactions and majorons.}---Another speculation holds that neutrinos could interact with each other more strongly than electroweak \cite{Berryman:2022hds}, to be differentiated from NSSI which often assume modest flavor-dependent modifications of the low-energy four-fermion interaction that is typically invoked in the context of nonstandard refraction and its impact on flavor conversion. The hypothesis of secret interactions, in contrast, often involves a light mediator $\phi$ that itself might be produced in a SN core, depending on its mass. If the interaction violates lepton number, $\phi$ is often called a majoron, and then decays of the type $\phi\to\nu\nu$ and other lepton-number violating processes can occur. 

The literature on majorons with mass is intertwined with secret neutrino interactions, and also with the question of internal SN deleptonization mentioned earlier. A simple SN~1987A cooling argument suggests that $g m_\phi<0.8\times10^{-8}$~MeV with $g$ the dimensionless Yukawa coupling and $m_\phi$ the majoron mass \cite{Heurtier:2016otg, Fiorillo:2022cdq}.  For such couplings, coalescence of highly degenerate neutrinos of the form $\nu_e\nu_e\to\phi$ and subsequent decays---outside of the SN---into all flavors through $\phi\to\nu\nu$ or  $\phi\to\bar\nu\bar\nu$, would lead to a 100-MeV-range detectable $\overline\nu_e$ flux \cite{Akita:2022etk, Akita:2023iwq, Fiorillo:2022cdq}. The absence of such events in the SN~1987A signal improves the cooling bound by an order of magnitude to $g m_\phi\lesssim0.8\times10^{-9}$~MeV, suggesting that such particles are not important for SN physics itself. 

The next Galactic SN could improve the constraints, and more intriguingly, could provide a positive detection \cite{Akita:2022etk}. This would allow one to reconstruct the majoron mass and coupling structure from the properties of the neutrino burst \cite{Telalovic:2024cot}.

{\bf Neutrino fluid and neutrino fireball.}---An intriguing question is what would happen if SN neutrinos interacted so strongly with each other that the neutrino-neutrino scattering rate would far exceed neutrino-matter interactions. There are no experimental tests of such speculations. In contrast to an early finding~\cite{Manohar:1987ec}, this effect would not strongly affect the streaming behavior from a SN core \cite{Dicus:1988jh}. Recently, this topic was revived with a new claim that the burst duration could be strongly affected by secret neutrino interactions~\cite{Chang:2022aas}; however, this was later falsified, as the resulting relativistic neutrino fluid produces a neutrino fireball, with an overall behavior surprisingly similar to the standard case \cite{Fiorillo:2023ytr, Fiorillo:2023cas}. Similar to lepton-number violation discussed earlier, even dramatic modifications of neutrino physics need not necessarily imply dramatic modifications of SN physics.

{\bf Sterile neutrinos.}---Very popular extensions of the neutrino sector invoke additional neutrino flavors that are sterile (inert) with regard to electroweak interactions, yet participate in mass--mixing \cite{Abazajian:2012ys, Boyarsky:2009ix, Giunti:2019aiy, Dasgupta:2021ies, Acero:2022wqg, Drewes:2013gca}---see a dedicated
chapter of this {\em Encyclopedia}. These are usually different from possible sterile Dirac partners of active species. Over the years, sterile neutrinos have been often invoked to explain various experimental anomalies, but also as dark matter candidates, notably with multi-keV masses. There could be one or more such species that dominantly mix with one or more of the active flavors, and their masses are unknown, so the phenomenological parameter space is huge. In addition, they can interact through magnetic dipole moments and decay radiatively as $\nu_s\to\nu_a+\gamma$. In a SN, they can carry away or transport energy and lepton number, help or hinder the explosion, modify nucleosynthesis, or become visible through decays outside of the SN. A concise synthesis of these different results is hard, so we merely refer to the large body of literature~\cite{Kainulainen:1990bn, Peltoniemi:1991ed, Mukhopadhyaya:1992dg, Raffelt:1992bs, Shi:1993ee, Pastor:1994nx, Kolb:1996pa, Nunokawa:1997ct, Sahu:1998jh, Caldwell:1999zk, Dolgov:2000jw, Fetter:2002xx, Keranen:2004rg, Beun:2006ka, Hidaka:2006sg, Hidaka:2007se, Keranen:2007ga, Choubey:2007ga, Fuller:2008erj, Kishimoto:2011mw, Raffelt:2011nc, Tamborra:2011is, Wu:2013gxa, Esmaili:2014gya, Warren:2014qza, Warren:2016slz, Arguelles:2016uwb, Yudin:2016zqp, Franarin:2017jnd, Jeong:2018yts, Mastrototaro:2019vug, Syvolap:2019dat, Suliga:2019bsq, Suliga:2020vpz, Tang:2020pkp, Chen:2022kal, Carenza:2023old, Chauhan:2023sci, Ray:2023gtu, Chauhan:2024nfa, Mori:2024vrf, Panda:2024avc, Ray:2024jeu, Chauhan:2025mnn, Mori:2025lky}.

{\bf Gravitons.}---Besides modifications of neutrino physics, other feebly interacting particles can affect SNe, and the most feebly interacting of all conceivable particles are gravitons $g$, the plausible but still hypothetical quanta of gravitational waves. Among other processes, they can be produced by nucleon bremsstrahlung ($NN\to NNg$) with a rate first estimated by Weinberg \cite{Weinberg:1972kfs}. In every $NN$ collision, the probability for emitting a thermal graviton with $E_g\sim T$ is very approximately $p_g\sim G_{\rm N}m_N^2 v_N^4$, where $G_{\rm N}$ is Newton's constant, $m_N$ the nucleon mass, and $v_N$ a typical thermal nucleon velocity. The resulting energy loss rate is negligible. 

It can become large in theories with large extra dimensions that predict a Kaluza-Klein (KK) tower of gravitons, i.e., gravitons also propagate in the higher-dimensional space and appear phenomenologically as if there were a dense multiplicity of gravitons with different masses, corresponding to their momenta in the hidden dimensions. The usual  SN~1987A cooling argument then provides restrictive limits on the size of the compact extra dimensions \cite{Cullen:1999hc, Hanhart:2000er}. Besides axion limits, these are the most-often cited SN~1987A cooling constraints on new particles. In addition, KK gravitons, which appear like particles with mass, can gravitationally accumulate around the PNS, form a halo around every NS, and can radiatively decay, leading to additional---but more model dependent---constraints \cite{Hannestad:2003yd}. After being forgotten for many years, this subject has recently enjoyed a revival of interest \cite{Law-Smith:2023czn, Im:2024jqx, Anchordoqui:2025nmb} in the context of the swampland program of String Theory \cite{Montero:2022prj}.

{\bf Axions.}---These pseudoscalar bosons are particularly well motivated  \cite{DiLuzio:2020wdo}, notably as wave-like dark matter, with a recent explosion of experimental search strategies, and therefore steady interest in astrophysical bounds \cite{Caputo:2024oqc, Carenza:2024ehj}. Axions derive from QCD and therefore generically couple to nucleons, quite generically to photons, and optionally to leptons \cite{GrillidiCortona:2015jxo, Springmann:2024mjp}. They are emitted from a SN core most efficiently by $NN$ bremsstrahlung, although pionic processes were also considered over the years, but for them, the nuclear-physics uncertainties are much larger. (A comprehensive discussion of the uncertainties, as well as a revised estimate of the axion emissivity, is given in Ref.~\cite{Fiorillo:2025gnd}.) SN~1987A axion bounds or detection forecasts were derived by many authors \cite{Raffelt:1987yt, Turner:1987by, Ellis:1987pk, Mayle:1987as, Ericson:1988wr, Mayle:1989yx, Brinkmann:1988vi, Burrows:1988ah, Burrows:1990pk, Turner:1991ax, Raffelt:1993ix, Janka:1995ir, Keil:1996ju, Hanhart:2000ae, Fischer:2016cyd, Chang:2018rso, Carenza:2019pxu, Carenza:2020cis, Fischer:2021jfm, Choi:2021ign, Vonk:2022tho, Ho:2022oaw, Betranhandy:2022bvr, Lucente:2022vuo, Li:2023thv, Lella:2023bfb, Springmann:2024ret, Mori:2025cqf, Fiorillo:2025gnd}, a vast literature that testifies to the broad community interest in axions over many years. The cooling bounds are roughly $g\lesssim 1\times10^{-9}$ for the dimensionless Yukawa coupling to nucleons, corresponding to an axion mass limit of $m_a\lesssim 13$~meV, the exact result somewhat varying with different SN assumptions and nuclear-physics input.

{\bf Axion-like particles (ALPs).}---Meanwhile, the idea of axions has been broadened to axion-like particles. This term sometimes refers to a hypothetical pseudoscalar with arbitrary mass, the same two-photon coupling vertex, and no other interaction. Sometimes it is taken to mean any such particle that is not a QCD axion, with various assumptions about interactions and mass. Concerning the impact on SN  questions, there is a vast literature studying ALPs coupling selectively to individual species, such as to electrons~\cite{Carenza:2021pcm, Fiorillo:2025sln}, muons~\cite{Croon:2020lrf,Bollig:2020xdr, Caputo:2021rux}, or photons alone \cite{Masso:1995tw,Jaeckel:2017tud,Caputo:2022mah,Diamond:2023scc,Fiorillo:2025yzf,Takata:2025lyu}.

{\bf Sub-GeV dark-matter particles.}---New sub-GeV particles might also act as dark-matter candidates, a topic that commands growing interest. A large variety of models can realize this phenomenology, including millicharged particles (MCPs) which can either couple directly to the SM photon with a small charge, or indirectly via a dark photon with a mixing to the SM photon \cite{Galison:1983pa, Holdom:1986eq, Dobroliubov:1989mr, Boehm:2003hm}. It is impossible to give justice to the large variety of models, so we focus instead on their possible impact on SN physics. Generally, these new particles can be constrained by the SN~1987A cooling argument~\cite{Davidson:2000hf, Knapen:2017xzo,Krnjaic:2015mbs,Chang:2018rso, DeRocco:2019jti, Manzari:2023gkt, Cappiello:2025tws}, which usually reaches below the direct detection constraints, but is limited at large couplings by trapping. On the other hand, these works all neglect the impact of self-interactions among dark particles; if their coupling to the dark photon---or any analogous mediator to the SM sector---is sufficiently large, this assumption breaks down. The new particles are then emitted as a fluid, rather than as kinetically streaming particles, which makes their outflow hydrodynamical and significantly affects the constraints, especially in the trapping regime~\cite{Fiorillo:2024upk}.

{\bf Cosmic diffuse SN flux of feebly interacting particles.}---If SNe emit large fluxes of axions or other FIPs, there will be cosmic diffuse fluxes in analogy to the DSNB, e.g., the DSAB for axions \cite{Raffelt:2011ft, Eby:2024mhd}, or analogous for ALPs \cite{Calore:2020tjw, Calore:2021hhn, Alonso-Gonzalez:2024ems, Candon:2025fnb} that, in principle, could be detected, or at least allows one to set constraint from the absence of photons from decays or magnetically induced~conversion.

{\bf Dark Photons.}---Another popular class of new particles are dark photons \cite{Caputo:2021eaa}, also called hidden photons, which are new vector-bosons with mass that mix with ordinary photons. They are often discussed as dark matter candidates in the multi-keV mass range. Astrophysical and other constraints are summarized in the exclusion plot of Ciaran O'Hare \cite{OHare}, which however ends at a mass of 100~keV. For larger masses, up to a few 100~MeV, the main constraints derive from SN physics \cite{Dent:2012mx, Kazanas:2014mca, Rrapaj:2015wgs, Chang:2016ntp, Caputo:2025avc}.

{\bf Shock heating and cooling.}---New particles, in analogy to standard neutrinos, can transport energy and contribute to shock rejuvenation or deposit energy directly in the stellar envelope, either way contributing to the explosion energy. This was perhaps the earliest idea about the impact of new particles on core-collapse SNe \cite{Falk:1978kf}, and low-energy SNe provide the most sensitive test \cite{Caputo:2022mah}. In special cases, the opposite is also possible: energy can be drained precisely from the gain region, where neutrino energy deposition should reheat the stalled shock. One scenario is the conversion between plasmons and dark photons, when the dark photon mass matches the plasma frequency of 100--400~keV in this region, leading to new constraints on dark-photon parameters~\cite{Caputo:2025aac}.

{\bf Trapping limit.}---New particles produced in SNe were often assumed to stream freely and simply drain energy because the interactions were assumed to be very feeble and the masses very small. However, particles can be so heavy that they can be produced, but either stay gravitationally trapped, decay quickly, or scatter so frequently that they contribute to energy transfer between different SN regions or between the PNS and the stellar envelope. The interaction of FIPs can be feeble by experimental standards, yet so strong that they have no impact on SN physics because their mean free path is short compared to that of neutrinos and their efficiency at energy transfer correspondingly smaller. 

In other words, usually there is a ceiling to the coupling strength excluded by SN physics, which can be difficult to identify. For axions, a pioneering treatment was provided by Burrows, Ressell and Turner \cite{Burrows:1990pk}, in this case picturing axions to be emitted from an axion-sphere in analogy to the neutrino sphere. For feebly interacting bosons, the theory of radiative transfer was recently studied \cite{Caputo:2022rca}. In the trapping regime, the energy transport from the PNS to the stellar material by means of new particles becomes diffusive, with an effective thermal conductivity; the corresponding rate of PNS cooling and energy deposition in the stellar material is obtained in Ref.~\cite{Fiorillo:2025yzf}. 

As a rule, the impact of a FIP on a star in general, or a SN core in particular, occurs when the mean free path is comparable to the geometric dimension. Being trapped alone does not make a FIP harmless, it has to be ``more trapped than neutrinos'' to become a perturbative effect instead of a dominating one.

{\bf Gamma rays from particle decays or magnetically induced conversions.}---FIPs, after leaving the SN, might decay or convert to gamma rays in intervening magnetic fields. At the time of SN~1987A, the Solar Maximum Mission (SMM) satellite was online, hosting a gamma-ray detector with sensitivity up to 100~MeV, that famously saw no feature in the background rate at the instant of SN~1987A, which was interpreted as a limit on neutrino radiative decays \cite{Chupp:1989kx, Kolb:1988pe, Oberauer:1993yr}. Later, the nonobservation with the Pioneer Venus Orbiter (PVO) were also used \cite{Jaffe:1995sw}. The radiative decays of FIPs can lead to further constraints, due to the nonobservation of high-energy gamma rays from SNe \cite{Jaeckel:2017tud, Meyer:2020vzy, Lella:2022uwi, Muller:2023vjm, Lella:2024dmx}, although for sufficiently large  couplings the photons produced in the decay are so compact that they form a fireball through gamma-gamma interactions, reprocessing their energy to the X-ray region \cite{Diamond:2023scc}; a similar fireball could be formed in NSMs, leading to competitive bounds from the merger GW170817 \cite{Diamond:2023cto}.
 
For particularly small masses ($m_a\lesssim10^{-9}$~eV), ALPs can coherently convert to photons in the Galactic $B$--field, and then the SMM limit provides constraints on the ALP-photon coupling strength of $g_{a\gamma}< 5\times10^{-12}~{\rm GeV}^{-1}$ \cite{Brockway:1996yr, Grifols:1996id, Payez:2014xsa, Hoof:2022xbe, Fiorillo:2025gnd}, a constraint shown in the context of other $B$--field conversion limits in O'Hare's famous compilation \cite{OHare}. In principle, the $B$-field of the progenitor star, which survives after collapse until the actual explosion, allows one to extend such constraints to larger masses \cite{Manzari:2024jns}, and with a satellite like Fermi-LAT---and for optimistic assumptions about the next Galactic SN---one may even become sensitive to QCD axions \cite{Manzari:2024jns, Fiorillo:2025gnd, Candon:2025sdm}, whereas the perspective from a NS merger is less promising \cite{Fiorillo:2025gnd, Lecce:2025dbz}. 

After an early warning of the next very nearby SN, one may direct the future IAXO helioscope toward the SN before it collapses and observe QCD axions by their conversion in the experimental magnet \cite{Ge:2020zww, Carenza:2025uib}. Of course, a detailed gamma-ray signal in coincidence with a future Galactic SN neutrino burst may allow one to probe both the ALP properties and the nuclear medium deep in the SN core \cite{Calore:2023srn, Lella:2024hfk}. 

While ALPs are purely speculative today, if they were detected, naturally they would become a new astrophysical probe, providing yet another multi-messenger channel.

\section{Neutrinos from the next Galactic supernova}
\label{sec:Next-Galactic}
\subsection{Galactic supernova rate}
\label{sec:SNrate}

\noindent After the enticing SN~1987A neutrino observations, a high-statistics measurement from the next nearby core collapse would quantitatively test the SN paradigm and provide new astrophysical and particle-physics information. When and where will it occur? Many operational and future large detectors have SN detection capabilities, but even the largest foreseen instrument, the Hyper-Kamiokande water Cherenkov detector that is under construction in Japan, has limited range. A SN in our large neighboring galaxy Andromeda (distance 765~kpc) would produce a few tens of events, comparable to the historical SN~1987A signal, and anyway, SNe in Andromeda are rare---the last one occurred in 1885 and was of type Ia (not core collapse). A high-statistics observation is therefore limited to the Milky Way and its satellites in the local group. A typical distance would be around 10~kpc, roughly corresponding to the Galactic-center distance of 8~kpc, but with a broad distribution reaching to around 20~kpc \cite{Adams:2013ana, Healy:2023ovi}, and as close as the nearest red supergiants at a few 100~pc (several hundred light years) \cite{Kato:2020hlc, Mukhopadhyay:2020ubs, ListRSG}, with Betelgeuse an often-discussed example. Expectations for neutrino signals are typically normalized to a distance of 10~kpc, but the true signal could be much stronger or weaker.

The Galactic core-collapse rate is estimated to be 1--3 cty$^{-1}$ (per century), with a recent best-fit value of $1.63\pm0.46$ \cite{Rozwadowska:2020nab}, which however draws from rather heterogeneous sources of information. This result is in line with a much older estimate of 1.7--2.8 cty$^{-1}$ \cite{Tammann:1994ev}, or with an estimate $1.9\pm1.1$ cty$^{-1}$ from radioactive $^{26}$Al (half life $7.2\times10^5$ yr) in the Galaxy that is produced by core-collapse SNe \cite{Diehl:2006cf}. The most direct constraint derives from the non-observation of any neutrino burst since 30~June~1980, when the Baksan Neutrino Observatory turned on~\cite{Alekseev:1993dy, Novoseltsev:2022lmd}, that is still running. Since that time, many detectors with SN sensitivity have come and gone and several very large ones are operating today. While there is no public record of the on and off times of all detectors since 1980, most of the time, if not always, one or more detectors must have been active. One can conclude with confidence that no Galactic SN has occurred in these 45 years, implying a 90\% CL upper limit of fewer than 5~cty$^{-1}$. The next nearby SN can also occur in the local group of galaxies, similar to SN~1987A in the LMC at a distance of 50~kpc, but the expected rate is smaller than for the Milky Way, although it is yet more uncertain \cite{Rozwadowska:2020nab}. 

The next nearby SN will be a once-in-a-lifetime opportunity, so it is reassuring that many large detectors are active or are under construction, with a long-term observation perspective for decades to come.

\subsection{Neutrino observatories}
\label{sec:nuobservatories}

\noindent Many neutrino detectors are operational world wide or under construction that can detect SN neutrinos (see Table~\ref{tab:detection-rates} and the SNEWS White Paper \cite{SNEWS:2020tbu}). Usually IBD ($\bar\nu_e p\to n e^+$) is the main detection channel that was discussed around Eq.~\eqref{eq:sigma_CC}. One observes the Cherenkov or scintillation light produced by the final-state positron, and sometimes in addition the neutron or its decay products or the gamma rays from $e^+e^-$ annihilation. The largest observatory of this kind is the water-Cherenkov detector Super-Kamiokande \cite{Super-Kamiokande:2016kji, Super-Kamiokande:2022dsn} that has operated since 1996 with several upgrades. In its most recent configuration, the pure water has been loaded with gadolinium \cite{Beacom:2003nk, Super-Kamiokande:2024kcb}, an efficient neutron absorber, a method that helps to identify particularly low-energy IBD events and is especially useful for DSNB detection. The signal from elastic electron scattering is subdominant, but reveals the SN direction, and is sensitive to all species. Another is the JUNO scintillator observatory  \cite{JUNO:2015zny, JUNO:2023dnp} that has recently (August 2025) begun taking data. Both would observe thousands of events for a SN at a distance of 10~kpc (Table~\ref{tab:detection-rates}). A yet larger water-Cherenkov detector Hyper-Kamiokande is under construction \cite{Hyper-Kamiokande:2018ofw, Hyper-Kamiokande:2021frf} with operation foreseen in or after 2028. All of these instruments, of course, have another main purpose, in particular to study neutrino flavor oscillations using reactor neutrinos (JUNO) or a neutrino beam (Super and Hyper-Kamiokande).  Beyond these main motivations, they have an extended portfolio of diverse physics goals. One detector built specifically for a Galactic SN watch was the 126~t LSD detector \cite{Dadykin:1987ek, Aglietta:1987it} that operated 1984--1999 in the Mont Blanc tunnel, but was too small to detect the SN~1987A burst in the LMC (see Sec.~\ref{sec:SN1987Adetection}). Another is the Large Volume Detector (LVD), running since 1992 in the Gran Sasso, that uses 1000~t liquid scintillator~\cite{LVD:2014uzr, Vigorito:2023}.

\begin{figure*}[t]
\vskip6pt
	\centering
    \hbox to\textwidth{\includegraphics[height=5.8cm]{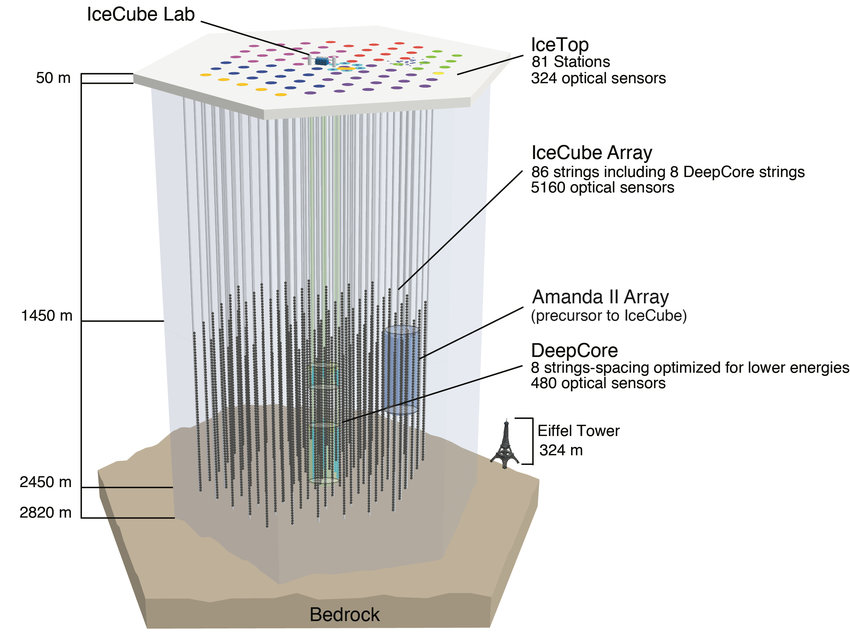}
    \hfill
    \includegraphics[height=5.8cm]{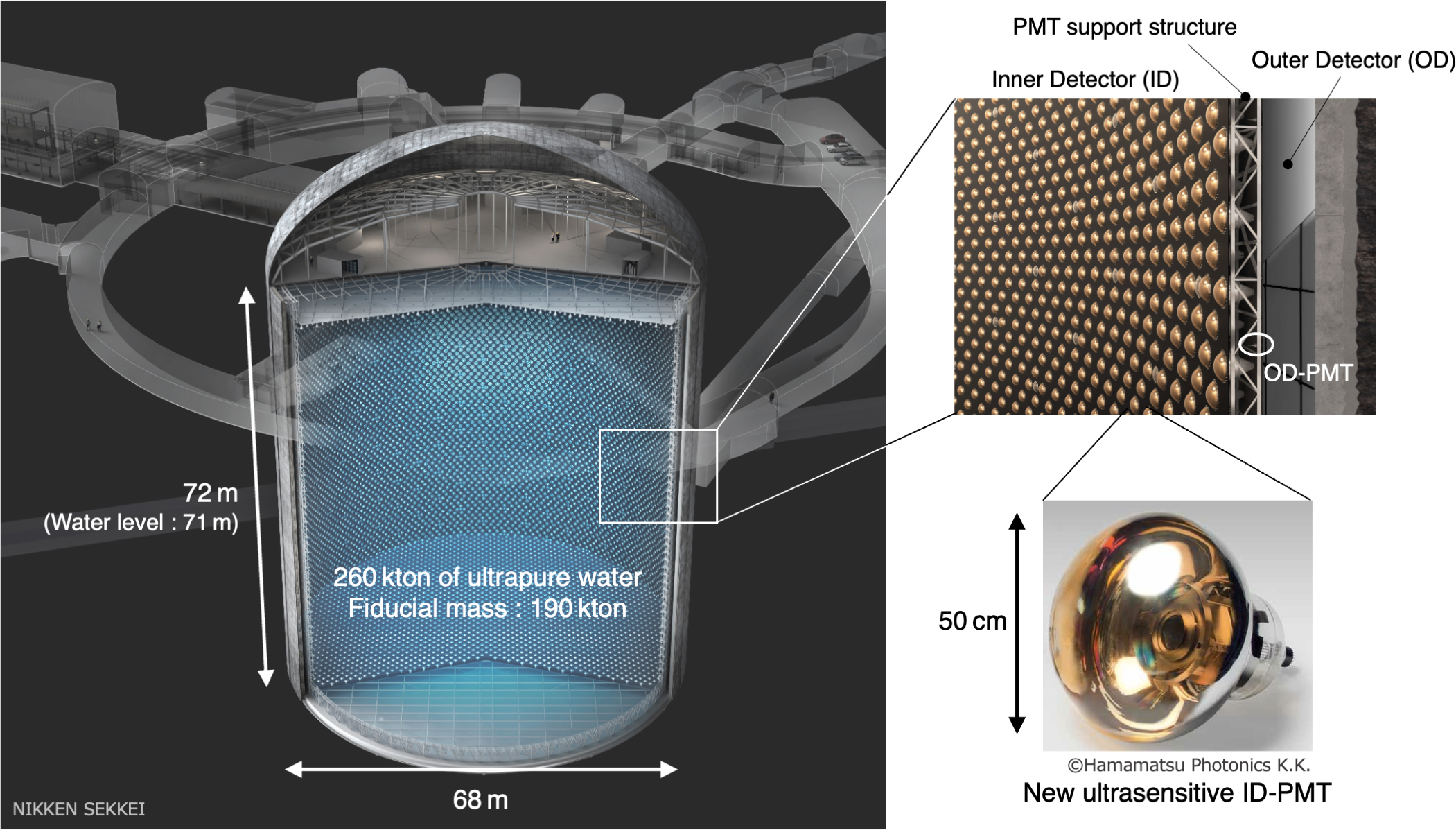}}
	\caption{The largest observatories for SN neutrinos. {\bf Left.} IceCube detector at the South Pole, primarily for high-energy neutrinos, but with fantastic sensitivity to a low-energy SN burst from ``correlated noise'' of its 5000 photomultiplier tubes (image \href{https://en.wikipedia.org/wiki/IceCube_Neutrino_Observatory}{Wikipedia}, created by K.~Andeen and M.~Plum for the IceCube Collaboration, licensed under \href{https://creativecommons.org/licenses/by/4.0/}{CC BY 4.0}). {\bf Right.} Hyper-Kamiokande under construction in Japan. With 220~kt fiducial volume it is the largest low-energy water Cherenkov detector ever built, foreseen to take up operation in 2028. (Image from \href{https://www-sk.icrr.u-tokyo.ac.jp/en/hk/}{Hyper-K Homepage} with permission, \copyright~Kamioka Observatory, ICRR, The University of Tokyo.)}
	\label{fig:Large-Detectors}
\end{figure*}

Besides Super-Kamiokande and JUNO, there are other truly large detectors, notably the high-energy neutrino telescope IceCube at the South Pole (Fig.~\ref{fig:Large-Detectors}) \cite{Abbasi:2011, IceCube:2023ogt}. It has 5000 optical modules (OMs) frozen in Antarctic ice over a volume of 1~km$^3$. The low-energy IBD products cannot be seen on an event-by-event basis, but can be identified by the overall increasing noise level from random photons being picked up when the ice flickers with Cherenkov light caused by a SN neutrino burst. The noise rate of a given OM is around 300~Hz and largely caused by radioactive impurities in the glass. The correlated noise from 5000~OMs from a SN actually provides superior time structure of the signal and could reveal fast time variations caused by hydrodynamical instabilities (Fig.~\ref{fig:SASI}). The same principle also works in analogous sea-water detectors, notably KM3NET \cite{KM3NeT:2021moe} that is under construction in the Mediterranean, although the large concentration of radioactive potassium $^{40}$K causes a large additional background of some 7~kHz per photo multiplier tube, and bio-luminescence is another background, all of which can be partly mitigated by coincidence requirements.

Detection channels other than $\overline\nu_ep\to ne^+$ or $\nu e\to e\nu$ open with other target materials. Most noteworthy are the liquid Argon time projection chambers for the DUNE long baseline experiment that is currently under construction in the US, where \hbox{$\nu_e+{}^{40}{\rm Ar}\to {}^{40}{\rm K}^*+e^-$} is most important for SN detection \cite{DUNE:2024ptd} that could help differentiate between the $\overline\nu_e$ and $\nu_e$ flux spectra and flavor conversion the neutrino vs.\ antineutrino sector. In liquid scintillator detectors, elastic scattering on protons would be an interesting if difficult-to-measure new channel \cite{Beacom:2002hs}. On the other hand, in dark-matter detectors with heavy nuclei as targets, in particular xenon, Coherent Elastic Neutrino-Nucleus Scattering (CE$\nu$NS) is sensitive to any $\nu$ or $\overline\nu$ flavor; ongoing and future large dark-matter detectors are beginning to have significant SN sensitivity \cite{Lang:2016zhv}. Yet other ideas include using archaeological lead \cite{Pattavina:2020cqc} or mineral deposits \cite{Baum:2023cct, Hirose:2025jht}.

Many of the active observatories are connected to the Supernova Early Warning System (SNEWS). Their White Paper and Webpage \cite{SNEWS:2020tbu} provide an excellent overview of neutrino detectors and physics opportunities. SNEWS collects tentative signals from different detectors and issue an alert to the community that a SN is imminent---the neutrinos arrive a few hours before the actual visible explosion. Depending on the actual observations and their relative timing, some advance directional information may also become available from neutrinos alone \cite{Beacom:1998fj, Tomas:2003xn, Fischer:2015oma, Brdar:2018zds, Hansen:2019giq, Linzer:2019swe, Coleiro:2020vyj, Sarfati:2021vym, DUNE:2024ptd}. For a particularly nearby progenitor, such as the often-discussed red supergiant Betelgeuse, even neutrinos from the pre-SN evolution might be recognized as a harbinger of imminent stellar collapse~\cite{Odrzywolek:2010zz, Kato:2020hlc}.

\subsection{What can we learn?}

\noindent Should we observe the neutrino signal from a collapsing star in our Galaxy, the scientific harvest would be immense \cite{Scholberg:2012id, Adams:2013ana, Mirizzi:2015eza, Horiuchi:2018ofe, Volpe:2023met, Volpe:2024zrw}. Exactly what we would learn, however, would depend on the details of the signal itself and on which detectors are operating at the time. If the observed features deviated significantly from the broad expectations outlined earlier, we might need to rewrite SN theory, despite its many convincing successes. Other crucial factors include  whether a coincident GW signal is detected, whether gamma-ray satellites are active, and what is known about the progenitor star. The impact of the information gathered from subsequent
multi-wavelength astronomical observations (possibly triggered by a SNEWS alert) can hardly be overstated. Finally, the statistical power of the neutrino signal, and the possibility of detecting even the progenitor neutrinos, will depend strongly on the~SN~distance.\looseness=1

\begin{table*}
\vskip8pt
	\caption{Event numbers for several detectors for three SN models (assumed distance 10~kpc), taken from the SNEWS White Paper and webpage \cite{SNEWS:2020tbu}. The models are s11.2c and s27.0c from \cite{Mirizzi:2015eza} that explode and form NS remnants, and s40 from~\cite{OConnor:2014sgn} that forms a black hole. We show the average of the event numbers for two cases of flavor conversion given in Ref.~\cite{SNEWS:2020tbu}. For KM3NET and IceCube, the detector mass is an effective value based on the number of events.\label{tab:detection-rates}}
    \footnotesize\sf
	\begin{tabular*}{\textwidth}{@{\extracolsep{\fill}}@{}lllllllll@{}}
			\toprule
			\multicolumn{1}{@{}l}{\TCH{Detector}} &
            \multicolumn{1}{l}{\TCH{Location}}&
            \multicolumn{1}{l}{\TCH{Status}}&
			\multicolumn{1}{l}{\TCH{Target}} &
			\multicolumn{1}{l}{\TCH{Mass [kt]}}&
            \multicolumn{1}{l}{\TCH{Species}} &
			\multicolumn{1}{l}{\TCH{$11.2\,M_\odot$}}&
			\multicolumn{1}{l}{\TCH{$27\,M_\odot$}}&
			\multicolumn{1}{l}{\TCH{$40\,M_\odot$ (BH)}}\\
			\midrule
			Super-K \cite{Super-Kamiokande:2016kji, Super-Kamiokande:2022dsn, Beacom:2003nk, Super-Kamiokande:2024kcb}
            &Japan        &ongoing&Water, Gd   &32      &$\overline\nu_e$&$4.1\times10^3$&$7.7\times10^3$&$6.3\times10^3$\\
            Hyper-K \cite{Hyper-Kamiokande:2018ofw, Hyper-Kamiokande:2021frf}     
            &Japan        &$>2028$&Water       &220     &$\overline\nu_e$&$2.8\times10^4$&$5.3\times10^4$&$4.3\times10^4$\\
            KM3NET       \cite{KM3NeT:2021moe}&Mediterranean&partial&Sea water   &(150)&$\overline\nu_e$&$1.9\times10^4$&$4.0\times10^4$&$5.1\times10^4$\\
            IceCube \cite{Abbasi:2011, IceCube:2023ogt}      
            &South Pole   &ongoing&Ice         &(2500)&$\overline\nu_e$&$3.3\times10^5$&$6.7\times10^5$&$7.3\times10^5$\\
            BUST \cite{Alekseev:1987ej, Alekseev:1988gp, Alekseev:1993dy, Novoseltsev:2022lmd}
            &Russia       &ongoing&Scintillator&0.24    &$\overline\nu_e$&$45$           &$85$           &$69$           \\
            SNO+ \cite{Caden:2024cgr}         
            &Canada       &ongoing&Scintillator&0.78    &$\overline\nu_e$&$150$          &$280$          &$230$          \\
            LVD \cite{LVD:2014uzr, Vigorito:2023}          
            &Italy        &ongoing&Scintillator&1       &$\overline\nu_e$&$190$          &$360$          &$290$          \\
            KamLAND \cite{KamLAND:2022sqb, KamLAND:2024uia}    
            &Japan        &ongoing&Scintillator&1       &$\overline\nu_e$&$190$          &$360$          &$290$          \\
            No$\nu$A \cite{NOvA:2020dll}     
            &USA          &ongoing&Scintillator&14      &$\overline\nu_e$&$2.0\times10^3$&$3.7\times10^3$&$3.0\times10^3$\\
            JUNO \cite{JUNO:2015zny,JUNO:2023dnp}
            &China        &ongoing &Scintillator&20      &$\overline\nu_e$&$3.8\times10^3$&$7.1\times10^3$&$5.8\times10^3$\\
            DUNE \cite{DUNE:2020zfm,DUNE:2024ptd}        &USA          &2030s&Argon  &40      &$\nu_e$    &$2.6\times10^3$&$5.4\times10^3$&$5.9\times10^3$\\
            &&&&\multicolumn{5}{l}{\TCH{Mass is in tons for the following small detectors}}\\
            DarkSide--20k \cite{DarkSide20k:2020ymr}
            &Italy        &$>2028$ &Argon      &39    &any $\nu$, $\overline\nu$&$120$ &$200$          &$110$          \\
            XENONnT \cite{Lang:2016zhv}      
            &Italy        &ongoing &Xenon      &6     &any $\nu$, $\overline\nu$&$39$  &$69$           &$44$           \\
            LZ  \cite{2018JInst..13C2024K}          
            &USA          &ongoing &Xenon      &7     &any $\nu$, $\overline\nu$&$33$  &$61$           &$42$           \\
            PandaX--4T \cite{Pang:2024bmg}   
            &China        &ongoing &Xenon      &4    &any $\nu$, $\overline\nu$&$26$  &$46$           &$29$           \\
			\bottomrule
	\end{tabular*}
	\vskip8pt
\end{table*}

From the particle-physics perspective, a smoking-gun signature for new particles could consist of unexpectedly high-energy neutrinos (perhaps in the 100-MeV range) coincident with the low-energy burst, indicating that radiation escaped from inside the SN core, perhaps through sterile neutrino production or majoron-type states. Likewise, a gamma-ray signal coincident with the neutrinos, or a sudden increase beginning with the neutrino burst, would prove the production and subsequent decay or conversion of axion-like particles or other feebly interacting particles. After an early warning of a nearby progenitor being about to collapse, a solar axion telescope such as (baby)IAXO might be pointing toward the SN and pick up axions. In general, any unusual signal in any detector, coincident with the neutrino (and possible GW) burst---but before the actual explosion---would probably indicate new physics. Any unusual timing features between different multi-messenger channels, or between different neutrino species, or depending on neutrino energy, would point to new propagation physics over large distances. Even in the absence of a clear signature for new physics, a high-statistics neutrino signal without unusual properties would allow one to constrain BSM physics along the lines of SN~1987A, but this time with superior~statistical~significance.\looseness=1

If there are no major surprises, the neutrinos, together with GWs (or their absence) and astronomical observations, will allow one---for the first time---to follow gravitational collapse, PNS evolution, and possibly BH formation in real time and with high statistics, possibly including fast SASI sloshing motions or other fast features. Current SN modeling is a parametric effort with regard to the nuclear equation of state and, to some degree, the neutrino opacities which depend on unmeasured nuclear correlation functions at finite temperature and in-medium coupling constants and interaction rates. PNS convection is also a major factor in shaping the neutrino signal. The underappreciated late-time signal at tens to hundreds of seconds after bounce would reveal information on fallback and/or the appearance of a nuclear phase transition. The different stages of the neutrino signal could be tested and the entire delayed explosion scenario, with all of its ramifications, will be on trial.

Neutrino flavor conversion may impact all emission phases, although would be especially relevant during the neutronization burst and accretion phase, when species dependence of the standard fluxes is largest. However, flavor conversion in itself is now an established phenomenon and probably all standard neutrino mass and mixing parameters will be eventually determined in laboratory experiments. This is a thoroughly different perspective compared to the time until 2012, when the discovery of a not-very-small neutrino mixing angle $\theta_{13}=8.5^\circ$ in the Double Chooz, Daya Bay, and RENO reactor-neutrino experiments opened the path toward the ongoing world-wide efforts in true three-flavor neutrino oscillation physics. Since that time, interest in SN neutrinos as a way to determine otherwise inaccessible mass--mixing parameters has~waned.

However, once all standard neutrino mass and mixing parameters are experimentally determined, the question can be turned around and one can use possible flavor-dependent features as a diagnostic tool on neutrino propagation in the dense SN environment or intervening space (``yesterday's sensation is today's calibration''). This perspective holds even if the next SN occurs tomorrow and such analysis has to wait until the completion of the experimental measurements. We are not sure if collective flavor evolution leads effectively to flavor equilibration---within the limits of conservation laws---near the decoupling region. We do not know if the flavor-dependent features during the accretion phase shown in the middle column of  Fig.~\ref{fig:NeutrinoSignal} exist or not. 

In other words, both for our understanding of the dense neutrino plasma and its impact on energy transport and explosion dynamics, it would be interesting to forecast the multi-detector, multi-location (relative to Earth matter effects), and multi-channel opportunities to answer this question without prior knowledge of the source spectra. Conversely, collective effects probably play no role during the onset period, the first tens of ms after core bounce, in connection with the prompt $\nu_e$ burst. Together with the future experimentally determined mass ordering, it will allow us to test the standard predictions from SN models together with MSW~conversion.

\section{Conclusion}
\label{sec:conclusion}

\noindent Neutrinos are the primary carriers of energy and lepton number in core-collapse SNe and other compact astrophysical transients unless new feebly interacting particles exist. After decades of dedicated effort, the long-standing Bethe--Wilson paradigm for SN explosions is strongly supported by numerical studies, with 3D simulations rapidly becoming the standard. Neutrino observations from the historical SN~1987A provided the first direct glimpse of this physics and motivated stringent particle-physics constraints on axions and other hypothetical feebly interacting particles. 

Yet, a high-statistics neutrino detection from a nearby SN remains crucial to move beyond this broad-brush confirmation toward a definitive test. Supernova modeling relies on simplifying assumptions---particularly regarding nuclear microphysics, the equation of state, and neutrino transport---so such a measurement would place the field on an entirely new footing of empirical certainty. At the same time, our theoretical grasp of collective neutrino flavor evolution is undergoing a transformation, driven by a new generation of researchers revisiting and revising long-standing ideas. 

While core-collapse SNe are remarkably diverse, each event unfolding with its own peculiarities, they share one unifying feature: the most elusive particles in the Universe orchestrate some of its most spectacular fireworks.

\section*{Acknowledgments}
\addcontentsline{toc}{section}{\protect\numberline{}Acknowledgments}

\noindent We thank Edoardo Vitagliano for useful comments on this manuscript. DFGF is supported by the Alexander von Humboldt Foundation (Germany). We acknowledge partial support by the German Research Foundation (DFG) through the Collaborative Research Centre ``Neutrinos and Dark Matter in Astro- and Particle Physics (NDM),'' Grant SFB-1258-283604770, and under Germany’s Excellence Strategy through the Cluster of Excellence ORIGINS EXC-2094-390783311. 

\section*{Glossary}
\addcontentsline{toc}{section}{\protect\numberline{}Glossary}

{\footnotesize
\begin{tabular}{@{}ll}
B & bethe (unit of energy for $10^{51}$ erg)\\
BH & Black Hole \\
BSM & Beyond Standard Model (particle physics)\\
BUST & Baksan Underground Scintillator Telescope \\
DSNB & Diffuse Supernova Neutrino Background \\
FIP & Feebly Interacting Particle \\
GW & Gravitational Wave \\
HST & Hubble Space Telescope \\
IBD & Inverse Beta Decay ($\overline\nu_e p \to n e^+$) \\
IMB & Irvine-Michigan-Brookhaven \\
Kam-II & Kamiokande II \\
LESA & Lepton-number Emission Self-sustained Asymmetry \\
LMC & Large Magellanic Cloud \\
LSD & Liquid Scintillator Detector\\
NS & Neutron Star\\
OM & Optical Module \\
pb & post bounce \\
PNS & Proto-Neutron Star \\
SASI & Standing Accretion Shock Instability \\
SM & Standard Model (of particle physics) \\
SNEWS & Supernova Early Warning System \\
SN & Supernova (plural SNe) \\
UT & Universal Time \\
WD & White Dwarf \\
$L_\odot$ & Solar luminosity $3.83\times10^{33}~{\rm erg}~{\rm s}^{-1}$ \\
$M_\odot$ & Solar mass $1.989\times10^{33}~{\rm g}$ \\
$m_{\rm eV}$ & $m/{\rm eV}$ (a particle mass $m$)
\end{tabular}
}

\addcontentsline{toc}{section}{\protect\numberline{}References}
\bibliographystyle{MyJHEP}
\bibliography{references}

\end{document}